\documentclass{aa}  
\bibpunct{(}{)}{;}{a}{}{,} % to follow the A&A style
\usepackage{graphicx}
\usepackage{lscape}
\usepackage{txfonts}
\usepackage{xcolor}

\begin{document}

\title{Properties of carbon stars in the Solar neighbourhood based on Gaia DR2
  astrometry\thanks{Table 1 is only available in electronic form at the CDS via anonymous ftp to cdsarc.u-strasbg.fr
    (130.79.128.5) or via http://cdsweb.u-strasbg.fr/cgi-bin/qcat?J/A+A/.}}

   \author{C. Abia\inst{1}
          \and
           P. de Laverny\inst{2}
          \and
          S. Cristallo\inst{3,4}
          \and
          G. Kordopatis\inst{2}
          \and
          O. Straniero\inst{3}
         }

   \institute{Dpto. F\'\i sica Te\'orica y del Cosmos, Universidad de Granada,
              E-18071 Granada, Spain
              \email{cabia@ugr.es}
              \and 
              Universit\'e C\^ote d’Azur, Observatoire de la C\^ote d’Azur, CNRS, Laboratoire Lagrange, France
              \and
              Istituto Nazionale di Astrofisica - Osservatorio Astronomico d’Abruzzo, Via Maggini snc, I-64100,
              Teramo, Italy
              \and            
            INFN -  Sezione di Perugia, via A. Pascoli, Perugia, Italy
          }

   \date{}

% \abstract{}{}{}{}{} 
% 5 {} token are mandatory
 
  \abstract
      {Stars evolving along the Asymptotic Giant Branch can become carbon-rich in the final part of their evolution.
        The detailed description of their spectra has led to the definition of several spectral types, namely: N, SC, J and
        R types. Up to now, differences among them have been partially established only on the basis of their chemical properties. }
      {An accurate determination of the luminosity function (LF) and kinematics  together with their chemical properties is extremely
        important for testing the reliability of theoretical models and establishing on a solid basis the stellar population membership
        of the different carbon star types.}
      {Using Gaia Data Release 2  (Gaia DR2) astrometry, we determine the LF and kinematic properties of  a sample of 210 carbon stars with
        different spectral types in the Solar neighbourhood, including some R-hot stars, with measured parallaxes better than $20\%$. Their
        spatial distribution and velocity components are also derived. Furthermore, the use of the infrared Wesenheit function allows us to
        identify the different spectral types in a Gaia-2MASS diagram.}
      {We find that the combined LF of N- and SC-type stars are consistent with a Gaussian distribution peaking at  M$_{\rm{bol}}\sim -5.2$ mag.
        The resulting LF however shows two tails at lower and higher luminosities more extended than those previously found, indicating that AGB
        carbon stars with Solar metallicity may reach M$_{\rm{bol}}\sim -6.0$ mag. This contrasts  with the narrower LF derived in Galactic carbon
        Miras from previous studies. We find that J-type stars are about half a magnitude fainter on average than N- and SC-type stars, while
        R-hot stars are half a magnitude brighter than previously found  although, in any case, fainter by several magnitudes than the rest of
        carbon types. Part of these differences are due to systematically lower parallaxes measured by Gaia DR2 with respect to  Hipparcos ones,
        in particular for sources with  parallax $\varpi < 1$ mas. The Galactic spatial distribution  and velocity components of the N-, SC- and
        J-type stars are very similar, while about $30\%$ of the R-hot stars in the sample are located at distances larger than $\sim 500$ pc
        from the Galactic Plane, and show a significant drift with respect to the local standard of rest.}
      {The LF derived for N- and SC-type in the Solar neighbourhood fully agrees with the expected luminosity of stars of 1.5-3 M$_\odot$ on
        the AGB. On a theoretical basis, the existence of an extended low luminosity tail would require a contribution of extrinsic low
        mass carbon stars, while the high luminosity one would imply that stars with mass up to $\sim 5$ M$_\odot$ may become carbon stars
        on the AGB. J-type stars not only differ significantly in their chemical composition with respect to the N- and SC-types but
        also in their LF, which reinforces the idea that these carbon stars belong to a dvifferent type whose origin is still
        unknown. The derived luminosities of R-hot stars make these stars unlikely to be in the red-clump as previously claimed. On
        the other hand, the derived spatial distribution and kinematic properties, together with their metallicity, indicate that
        most of the N-, SC- and J-type stars belong to the thin disc population, while a significant fraction of R-hot stars show
        characteristics compatible with the thick disc.}


   \maketitle
   
 % \authorrunning{C. Abia et al.}
 %\titlerunning{Carbon stars in the solar 
 % neighbourhood}

\section{Introduction}

After He exhaustion, low and intermediate mass stars ($0.8\leq$ M$/$M$_\odot\leq 8$) populate
the giant branch in the Hertzsprung–Russell diagram for a second
time. Stars in this phase of stellar evolution are known as asymptotic giant branch (AGB) stars.
AGB stars are very important contributors ($>50\%$) to the mass ejected by all stars into the interstellar medium (ISM). Therefore,
they play a significant role in the chemical evolution of galaxies \citep[see, e.g.][]{wal97}. Furthermore, they trace intermediate
age stellar populations so that they can be used in studies of Galactic structure \citep[e.g.][]{col02,ren15,cap16,jav18}. All those
research topics rely on an appropriate estimation of the stellar parameters, and in particular on the mass. 

Probably the most important chemical peculiarity of AGB stars is that many of them are carbon rich, i.e. they show an abundance
ratio C/O$>1$ (by number) in their atmosphere. Since the overwhelming majority of stars are born with C/O$<1$, this carbon enrichment
must result from an {\it in situ} process that pollutes the envelope with fresh carbon produced in the interior
or, alternatively, from a transfer of carbon-rich material in a binary system. In the first case the stars are named {\it intrinsic}
carbon stars and, in the second case, {\it extrinsic} carbon stars. It is well established that quasiperiodic shell $^4$He burning occurs
during thermal pulses (TPs) on the AGB, inducing mixing episodes (the third dredge-up, TDU) that increase the atmospheric C/O ratio in
these red giants \citep[see, e.g.][]{ibe83,stra06,kar14}. This gives rise to the spectral sequence of M to MS to S to SC to C as the C/O
ratio increases in the envelope along the AGB phase. This increase in C/O is accompanied, in general, by an increasing s-process
overabundance. Thus, the MS, S, SC and C stars are heavy element rich AGB stars, which has been confirmed by many spectroscopic
studies \citep[see][among many others]{smi90,bus02,lam95,van98,van99,abi98,abi02}. The mass range for the formation of an AGB
carbon star is still, however, rather controversial. This is  mainly due to our limited modelling of the TDU episodes as well
as to our poor knowledge of the mass loss rate occurring during the AGB. Despite of this, there is an ample consensus that the
lower mass limit for the formation of a carbon star at solar metallicity is $\sim 1.5$ M$_\odot$, and that this limit is decreasing
with  decreasing  metallicity. The latter is due to the fact that metal-poor stars have less O in their envelopes (it is therefore
easier to rise the C/O above unity) and to the fact that the efficiency of the TDU increases at low metallicity. The upper mass
limit is, nevertheless, more uncertain. Theoretically, stars with M$\ga 3-4$ M$_\odot$ may burn hydrogen through the CNO by-cycle
at the base of the convective envelope (the so called hot bottom burning, HBB), avoiding the formation of an AGB carbon star and
substantially altering the CNO isotopic ratios, lithium and other light element abundances (e.g. F, Na) in the envelope. This
depends on the actual mass and metallicity of the star. However, these findings are  extremely dependent on the mixing treatment,
mass-loss and thermonuclear reaction rates adopted in the stellar modelling \citep[see][for detailed discussions]{lat03,stra06,kar14,ven15}.
Unfortunately, very few observational studies of massive AGB stars exist to constraint theoretical models in this sense \citep{mcs07,gar07,abi17},
not to mention the difficulty in the determination of accurate masses (luminosities) for these stars  \citep[see e.g.][among others]{loo97,fro98,loo99,mar99,zij06,gro07,pas19}.

Until recently, our knowledge of the luminous red giant carbon stars was limited to their spectral types, inaccurate radial
velocities, some uncertain proper motions, but detailed descriptions of their spectra. The last  of these included the
identification of key molecules
such as C$_2$, CH, and CN in the visual and infrared wavelengths and the recognition that some had enhanced lines of
heavy elements. There are several types of carbon stars classified spectroscopically, mainly depending on the intensity
of the molecular bands mentioned above and their 
effective temperature \citep{kee93, bar96, wal98}. During the past few decades, quantitative abundance and isotopic
ratio determination in carbon stars of all types has allowed to differentiate among them, as well as a better understanding
of their nucleosynthetic histories 
(sometimes affected by binarity) and evolutionary status. Among the red giant carbon stars four spectral types can be
distinguished \footnote{There are other types of less evolved carbon stars {\bf such} as the dwarf carbon stars, CH,
  Ba, and CEMP (Carbon Enhanced Metal Poor) stars. They are main-sequence or sub-giant stars and, although they show
  carbon enhancement, the C/O ratio does not necessarily exceed unity in the envelope. We do not study these stars here.}.
In what follows, we summarise their main properties; a more detailed and extended discussion can be found in \citet{jur91,bar96,wal97,wal98,kna01,ber02,abi03,zam09},
and references therein.

a) The normal N-type (or just C) are the most numerous ones. They are cool (T$_{eff}< 3500 $ K) and luminous objects ($\sim 10^4$ L$_\odot$).
Their spectra are very crowded showing intense carbon bearing molecular absorptions and, in particular, a strong flux depression at $\lambda \la 4000$ {\AA}.
By definition these stars show C/O$>1$ (although not much larger than unity) and most of them are enhanced in s-elements and
fluorine at typically near solar metallicity ([Fe/H]$\approx 0.0$)\footnote{We adopt here the usual notation [X/H]$=$ log (X/H)$_\star-$ log (X/H)$_\odot$,
  where (X/H)$_\star$ is the abundance by number of the element X in the corresponding star.}. The observed carbon, s-element and fluorine enhancement are
believed to be produced by
the recurrent mixing into the envelope of material exposed to He-burning through the TDU episodes, i.e. they are intrinsic
carbon stars. Due to their large  luminosities, they can be easily identified in the  galaxies of the Local Group \citep[see e.g.][]{row05,boy13,whi18}.

b) The stars of SC-type are characterised mainly by a C/O$\approx 1$, which makes their spectra less crowded by molecular
absorptions allowing the identification of a plethora of atomic lines. In the classical picture, these stars should correspond
to a very short period in the evolution along the AGB phase when the star is transformed from an M  (MS, S) (with C/O$<1$)
into a genuine carbon star (N-type with C/O$>1$) due to the TDU episodes. Their chemical properties are indistinguishable from
those observed in the N-type (or in the S-type stars, see e.g. \citet{van98,ney11}), although there is a hint indicating
that SC-type show higher $^{16}$O/$^{17}$O than the N-type \citep{abi7}. Since high $^{16}$O/$^{17}$O ratios is a characteristic
of the operation of HBB, if confirmed, these stars may be intermediate mass ($\ga 3-4$ M$_\odot$). They then might become C-rich
at the very end of their life because of the cessation of the HBB when their envelope mass has been significantly reduced
by strong mass-loss \citep[see e.g.][]{kar14}. Only a  dozen or so  of SC-type stars are identified in the Galaxy \citep{cat71},
and a few of them are among the most Li-rich stars ever found  \citep{abi99}.\footnote{ Very luminous Li-rich carbon stars have
  been also found in the Magellanic Clouds \citep{smi95}.}

c) The J-type carbon stars are easily recognised from the intensity of carbon bearing molecular bands formed with $^{13}$C atoms;
in fact their main chemical property is their very low $^{12}$C/$^{13}$C ratio, close to the CNO-cycle equilibrium value ($\sim 3.5$).
They show  C/O ratios in a range similar to that found in the N-type but a significant fraction ($80\%$) of them are Li-enhanced.
They do not  show  s-element nor fluorine enhancements \citep{abi00,abi15}. They are also Solar metallicity stars. These chemical
peculiarities clearly differ from those of N- and SC-type stars. Therefore, the location of J-type stars in the AGB spectral
sequence above is far from clear. It has been suggested that the mass transfer scenario is at the origin of their carbon enhancement
(i.e. they would be extrinsic carbon stars). In fact, in some of them, silicate emission has been detected in their infrared spectrum,
which usually is associated to the presence of a circumbinary disc around a binary system \citep[see e.g.][]{der07}.  However,
aside from the fact that no radial velocity variations have been  detected yet in any of these stars, it is not clear how their
Li enhancement can be explained in the mass transfer episode. Other scenarios have been proposed,  such as the mixing of fresh
carbon after a peculiar  He-flash induced by a rapidly rotating core \citep[e.g.][]{men76}, or the re-accretion of carbon-rich
nova ejecta on main-sequence companions to low-mass carbon-oxygen white-dwarfs \citep{sen13}. Despite their origin is  unknown,
they represent about $15\%$ of all Galactic carbon stars and they have also been identified in Local Group galaxies in a similar
fraction \citep{mor93}. 

d) Finally, R-hot (or early) carbon stars are the warmer (T$_{eff}\ga 3800$ K) objects and are easily identified spectroscopically
because of their less crowed spectra and weaker molecular absorptions as compared to the rest of the carbon star types\footnote{We
  do not discuss here the named R-cold (or late) carbon stars as it has been shown that they are indistinguishable from the N-type
  stars \citep{zam09}.}. There is not a genuine chemical property in these stars except that they do not show s-element enhancements,
their metallicity being also typically Solar \citep{domi84,zam09}. It is well established that these stars are much fainter than the
rest of the carbon types. Using Hipparcos parallaxes, \citet{kna01} placed them in the red-clump region on the H-R diagram with a
mean luminosity M$_K\approx -2.0$ mag, which is compatible with core He-burning stars. Further, most of these early R-type stars
are non-variable, and their infrared photometric properties show that they are not undergoing significant mass loss, as opposed
to the other three spectral types. As for the J-type stars, no radial velocity variations have been detected in these stars \citep{mac84},
which would discard the mass transfer scenario as an explanation of their carbon enhancement.  The favoured hypothesis so far is that the
carbon produced during the He-flash is mixed in some way to the surface. However, the general conclusion is that hydrostatic models do not
produce mixing at the He-flash, thus other scenarios have been explored. An anomalous He-flash after a red giant star's  merging  has been
suggested, but this scenario  has many difficulties to explain the observed chemical properties \citep{iza07,pie10}. Recently, \citet{zha13}
found that a high-mass helium white dwarf subducted into a low-core-mass red giant could produce a R-hot star with the observed chemical
properties. Furthermore, in this scenario J-type stars may represent a short and luminous stage in the evolution of a R-hot star. Nevertheless
considering that R-hot stars represent  $\sim 30\%$ \citep{ber02} of all luminous red giant carbon stars, there is a necessity to show
that this scenario can account for the observed statistics. 

As it follows from the above discussion, similarities and differences among the different spectral types have been made mainly on the
basis of their observed chemical pattern. However, the conclusions deduced from these abundance studies are frequently limited by the
uncertain determination of the mass, luminosity and kinematics properties of the stars. The Gaia
all-sky survey is changing this situation providing astrometric data with an unprecedented accuracy of all Galactic stellar populations.
In the Gaia DR2 release \citep{gai18}, the first Gaia catalogue of long period variables (LPVs) with G-band variability amplitudes larger
than 0.2 mag has been published \citep{mow18}. It contains 151\,761 candidates, among them thousands luminous carbon stars of variable
types Mira, irregular and semi-regular. The aim of the present study is to use the Gaia DR2 information available on a sample of already
known red giant carbon stars of different spectral types located in the Solar vicinity to derive their luminosity and kinematic properties.
These  quantities are discussed together with their chemical characteristics  to obtain a global picture of the different carbon star types
and to discern between the various types in terms of luminosities, masses, evolutionary stages and kinematics. 

The structure of the paper is as follows: in Sect. 2 we describe the adopted sample of carbon stars and study their spatial distribution
in the Solar neighbourhood. Sect. 3 describes the derivation of the luminosity distribution of the sample and its implications on the
possible masses and evolutionary status of the sample stars. In Sect.~4, we derive their kinematic properties. In  Sect. 5, we show how
the combination of Gaia and infrared photometry allows the identification of the different spectral types of carbon stars in the Gaia-2MASS
diagram, as it has been recently done for the LPV in the Large Magellanic Cloud. Finally, in Sect. 6 we summarise the main results of this
study.

%-----------------------------------------------------------------
\section{Stellar sample and spatial distribution}

Different samples of Galactic AGB stars, selected on the basis of their infrared properties, can be found in the literature, for
instance \citet{cla87,wil88,jur89,gro92}. Among these surveys the more extensive is that by \cite{cla87}, which focuses exclusively
on luminous carbon stars. This survey was  in fact used by \citet{bof93} and \citet{abi93} to study the frequency of the Li enrichment
among Galactic carbon stars. Furthermore, within this catalogue, there is valuable information on the chemical properties for many objects,
obtained from high-resolution spectroscopic studies (see references in Sect. 1). We decide, therefore, to base our study on the survey by \citet{cla87}.
This sample was drawn from the {\it Two Micron Sky Survey} \citep{neu69}, which is expected to be statistically complete for sources brighter
than $K\sim +3.0$ mag over the region $-33^\circ<\delta<+81^\circ$. This sample contains 214 Galactic carbon stars to which we have added some
additional carbon stars  with already well determined photospheric characteristics and, in particular,  chemical properties \citep[e.g.][and r
  eferences therein]{lam86,abi93,ohn96,abi98,abi00,abi02,vant07,abi15}. Spectral types were taken directly from
the SIMBAD\footnote{http://simbad.u-strasbg.fr/simbad/sim-fid} database, although a few stars were re-classified according to the
more detailed spectroscopic studies mentioned above.  
This initial sample was then filtered considering the quality of their Gaia DR2 parallaxes \citep{gai18}. For the present study,
only stars with good-quality DR2 parallaxes  ($\varpi$) are selected, that is, those matching the condition $\epsilon(\varpi)/\varpi\leq 0.20$.
We note, however, that the overwhelming majority of the stars ($\sim 85\%$) in our sample have Gaia DR2  parallax uncertainty $\leq 10\%$.
This condition ensures that the distance adopted from \citet{bai18} is close to the inverse of the parallax, and therefore has
little dependence on the adopted prior.  Applying the above parallax uncertainty criterion, our final sample consists of 10 stars
of SC-type, 22 of J-type, and 143 of N-type; $80\%$ of them having parallax uncertainty smaller than 10\%. Furthermore, for comparison
purposes, we added 10 O-rich AGB stars of M-type and 35 R-hot type carbon stars fulfilling the above mentioned parallax criterion,
the later ones selected from the magnitude-limited study of \citet{kna01}; some of them being analysed chemically by \citet{domi84} and \citet{zam09}.
We point out that several of the R-hot stars included in the study of \citet{kna01} were discarded here because they have been re-classified as CH,
Ba and/or CEMP stars or because their
spectral type appears rather doubtful in the literature\footnote{We note that the two micron sky survey is less sensitive to the
detection of R-type stars than to the other carbon star types \citep{cla87}.}. The few O-rich AGB stars were taken from the
study of \citet{gar07} and are expected  to be intermediate mass stars probably undergoing HBB because of their strong Li absorption
at $\lambda  6078 \AA$ (see below).

Except for the R-hot type, the vast majority of the stars in our sample are known or suspected variables. We have 28 Mira variables
among the C- and O-rich stars, the others being irregulars or semi-irregular variables. Only a couple of R-hot stars in the sample,
however, are classified as variable of R CrB type. Finally, from the chemical studies mentioned above, the overwhelming  majority of
the selected stars have near Solar metallicity ([Fe/H]$\sim 0.0\pm 0.3$).
As a consequence and unless explicitly mentioned, we will assume throughout this paper a Solar metallicity for all of the stars in
our sample.

\begin{figure}
\centering
\includegraphics[width=9.0cm]{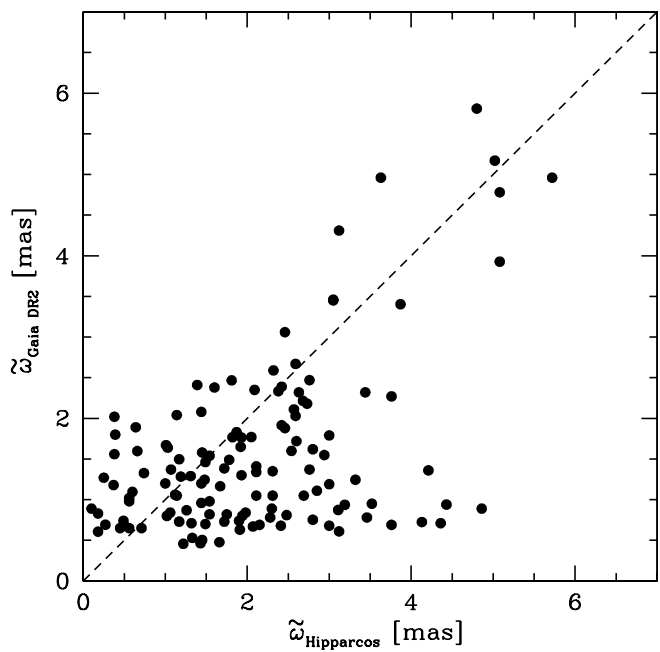}
\caption{Comparison between the Hipparcos and the Gaia DR2 parallaxes
for 139 stars in common with our sample of Galactic carbon stars.
Typical parallax errors are of the order of $50\%$ (or even large for Hipparcos and smaller
than $15\%$ for Gaia. Error bars are omitted in the plot for clarity. Clearly Hipparcos parallaxes
are systematically larger for sources with Gaia
$\varpi \leq 2$ mas. See text for details.}
\end{figure}

The  sample stars are listed in Table~1 (col. 1), adopting preferentially their variable star designation \citep[from][hereafter GCVS]{sam04} or,
if not available, their
most used name in the literature (as checked with the SIMBAD database).
For the R-hot stars, we used instead their Hipparcos catalogue identification.
For all the sample stars, we then searched for their Gaia DR2 identification in the Gaia database web facility using the adopted target
names of Table~1 and recovered all the Gaia astrometric information available. We adopted the distance estimates and associated
uncertainties of \cite{bai18} for all of the sample. For the line-of-sight velocities, we first adopted the values given in \citet{men06}
for thirteen stars that rely on CO millimetre lines measurements. These estimations are more accurate than classical ones from  near
infrared (NIR) or optical spectral lines. Typical uncertainties reported in this work are of the order of a few km\,s$^{-1}$. For the other
stars, we first retrieved from CDS/SIMBAD all the available bibliographic heliocentric radial velocity measurements, V$_{\rm{rad}}$. When possible,
we favoured the values reported in the {\it Pulkovo Compilation of Radial Velocities} \citep{Gontcharov} that have a median accuracy of about $\pm 1$ km\,s$^{-1}$. If not present in this catalogue either, we adopted the V$_{\rm{rad}}$ measurement of the most recent reference found in SIMBAD. Finally, for the few stars without any reported V$_{\rm{rad}}$ in the above references, we adopted the Gaia DR2 V$_{\rm{rad}}$ \citep{Katz2019}. We decided not to adopt a-priory the Gaia DR2 V$_{\rm{rad}}$ for all our sample stars since the Gaia/RVS spectral domain is not optimal for deriving V$_{\rm{rad}}$ in cool carbon rich stars. Indeed, their spectra are crowded by molecular absorptions and the IR CaII triplet is usually not easily visible in such  stars \citep[see e.g.][]{bar96}. Moreover, as far as we know, the Gaia DR2 pipeline for V$_{\rm{rad}}$ derivation is not optimised for analysing (not a-priory known) carbon enhanced spectra.  This, however, does not concern the R-hot stars, for which, due to their warmer effective temperature (T$_{\rm{eff}}\geq 3800$ K), the IR CaII triplet can be easily identified and thus useful for radial velocity determinations. For these stars, we have taken directly the Gaia DR2 V$_{\rm{rad}}$ values \citep{Katz2019}.

The distances and V$_{\rm{rad}}$ values finally adopted in this study are reported in Table 1. These are the main parameters from which the luminosities and kinematic properties are derived.  To check the quality of the Gaia DR2 astrometric data for our sample stars,
we have looked at their Renormalised Unit Weight Error (RUWE).
This error coefficient can be used to identify possible non-single stars and/or possible problematic cases of the astrometric
solution \citep[see Gaia DR2 documentation, Sect.14.1.2 and][]{Lindegren18}. Typically, according to Gaia's documentation, poor astrometric fits have a RUWE parameter
larger than $\sim$1.4. We find that 70\%, 90\% and 94\% of
our sample stars have a RUWE factor less than 1.0, 1.2 and 1.4, respectively. 
We have also checked the correlations between the RUWE factor and the parallaxes or the proper motions. Whereas no correlation is seen with the parallaxes, a slight increase of the proper-motion error for the stars having RUWE greater than 1.2 is seen, increasing from $\sim 0.2$ mas yr$^{-1}$ to $\sim 0.4$ mas yr$^{-1}$. We note, however, that these errors still remain small  (and affect less than 10\% of the stars). We are therefore confident in the astrometric data adopted in this study. Nevertheless, we  note here that the Gaia DR2 procedure to compute the parallax contains colour- and magnitude-dependent terms. For the Hipparcos survey, \citet{pou03} and \cite{pla03} advocated the importance of the colour bias on the astrometry, particularly for very red objects. Because most stars in our sample show large colour variations within a cycle (they are variable stars), chromaticity corrections must be applied to each epoch data. On the contrary, Gaia DR2 parallaxes are determined assuming a constant mean colour and magnitude for each source \citep{lin18}. 
  Another phenomenon that may affect the parallax measurement of late-type stars, in particular its accuracy, is the possible displacement of the photometric centroid (photocentre) of the star with respect to the projected barycentre due to surface brightness asymmetries  (e.g. due to granulation- or convection-related surface variability). \citet{luw06} has shown that these effects may be considerable in red supergiants. Recently \citet{Chiavassa18} studied the limits in astrometric accuracy of Gaia induced by these surface brightness asymmetries on the basis of radiative hydrodynamic simulations of AGB stars. They conclude that the displacement can be a non-negligible fraction of the star radius R ($5-10\%$ of the corresponding stellar radius), accounting for a substantial part of the parallax error. The above issues, may affect the measured parallaxes  as well as their accuracy\footnote{ We do not expect the proper motions to be affected by this effect, as the signature on the sky over the time span of Gaia GDR2 observations allow to distinguish clearly the proper
    motion movement from the parallax effect.}. We cannot exclude, therefore, that the parallax for some of our selected stars could be incorrectly evaluated in Gaia DR2. Our conclusion, nevertheless, is that our results should not be affected significantly by possible astrometric problems since the RUWE parameters of our sample are in average quite small (median$=0.94$ and mean$=1.1)$. Obviously the increase of the parallax estimates over one-year cycles expected in Gaia DR3 would reduce the error in the mean parallax.

Figure 1 shows the comparison between  Hipparcos \citep[][for R-type stars]{van07,kna01} and Gaia DR2 parallaxes for 139 stars in common with our sample. From this figure, it is evident that Hipparcos parallaxes tend to be much larger than Gaia DR2 ones, particularly for $\varpi \la 1$ mas. For $\varpi \ga 1$ mas, Gaia DR2 and Hipparcos parallaxes
appear to be symmetrically distributed around the equal-parallax line. Globally, we find a mean difference of $\Delta\varpi=0.56 \pm 1.70$, in the sense Hipparcos minus Gaia DR2. Obviously this difference has a significant impact on the luminosities derived for our stars as compared to those based on Hipparcos parallaxes (see  Sect.~3).

Figure 2 shows the location of the sample stars onto  and above/below the Galactic plane (see also Table 1). Their Cartesian coordinates have been directly derived from the Gaia DR2 sky coordinates and adopted distances. 
The uncertainties on the individual X, Y, Z and R positions are estimated as the dispersion of these values for each star, over 500 Monte-Carlo realisations of the \citet{bai18}
line-of-sight  distances\footnote{We assumed a symmetric uncertainty derived as the average of $\rm (r_{est}-r_{low})$ and $\rm (r_{high}-r_{est})$, see \citet{bai18}
  for further details.}, assuming no uncertainty on the right ascension nor declination sky coordinates (see also Sect.~4). Median uncertainties are $\pm 12, 16, 13$ and $13$\,pc
for X, Y, Z and R, respectively.
%Typical uncertainty, computed as the dispersion  is of the order of $\pm 20$\,pc for  (X, Y, Z), and $\pm50$\,pc for the galactocentric distances (R). 
%Typical uncertainty is of the order of $\pm 20$\,pc for  (X, Y, Z), and $\pm50$\,pc for the galactocentric distances (R). 
%

From the left panel of this figure, it appears that the spatial distribution of the N-, J- and SC-type stars (in blue, red and green, respectively) is fairly uniform  within a radius of $\sim 1.5$ kpc from the Sun, in the region of the sky observed by the two micron sky survey. In fact, the apparent lack of these spectral type sources in the region with X$<0$ and Y$<0$ is simply due to the incomplete coverage of this survey: the southern declination limit being $-33^\circ$. A few N-type stars, however, seem to be located beyond this radius. Furthermore, for these spectral types, it can also be appreciated that there is no specific concentration of carbon stars toward either direction. This agrees with the conclusions of  \citet{cla87}. In contrast, our sample of R-hot stars (brown circles in Fig.~2) seem less homogeneously distributed around the Sun with, apparently, an overdensity around it. The left panel of Fig. 2 shows that carbon stars of N-, J- and SC-types are rather concentrated towards the Galactic plane with no appreciable difference between  the different spectral types in the height from the plane.
\citet{cla87} argued that the present sample of carbon stars 
should be complete within 1.5 kpc from the Sun on the hypothesis that all the carbon stars have
an absolute magnitude M$_K = - 8.1$ mag (we confirm this hypothesis here, see below) and considering that the limit of the two micron sky survey in $K$ is $\sim + 3.0$ mag. A comparison with the number of carbon stars detected in other similar surveys shows that this expectation is  fulfilled \citep[see][for details]{cla87}. On this basis, a crude fit to the z-coordinate distribution, excluding the R-hot stars, with an exponential function results to a scale-height  of $z_o=180\pm 20$ pc.
Considering only the N-type stars would slightly change this value.
We note that only two N-type stars, V CrB and RU Vir, are located well above the Galactic plane. 
We will see in the following sections that also their peculiar kinematics could point towards a possible thick disk (or halo) membership or even an extragalactic origin. Regardless of these two stars, the measured scale-height agrees well with the range 150-250 pc estimated by
\citet{cla87}. This scale-height can then be used to estimate the typical mass of carbon star progenitors by using tabulations for the scale-heights of main-sequence stars as a function of spectral class as, for example, in  \citet[][]{mil79}. This study showed
that stars with a mean scale-height in the range 160-190 pc have a mass
between 1.5 and 1.8 M$_\odot$. These values are fully consistent with theoretical determinations of the typical mass for an AGB carbon star \citep[see e.g.][]{stra06,kar14}. Contrary to N-, J- and SC-type stars, it is also evident from Fig.~2 that the R-hot carbon stars %are distributed in a wider distance range  above/below the Galactic plane. 
reach larger distances from the Galactic plane than the rest of carbon type stars.
In fact, they are distributed quite uniformly in $|z|$ within $0\leq |z|\leq 1$\,kpc, the average value being $|z|=510\pm 300$\,pc; only one star HIP 48329 is located at $|z|> 1$\,kpc (not shown in Fig. 2 for clarity, see Table 1). Since the Galactic stellar population scale-height is believed to be a function of age (mass) \citep[see e.g.][]{dov93}, this implies that R-hot stars are likely older and have lower masses than the N-, J- and SC-type stars. This also confirms the conclusion already reached in previous studies \citep{wal98,kna01,iza07,izz08,zam09}.

\begin{figure*}
   \centering
   \includegraphics[width=17cm]{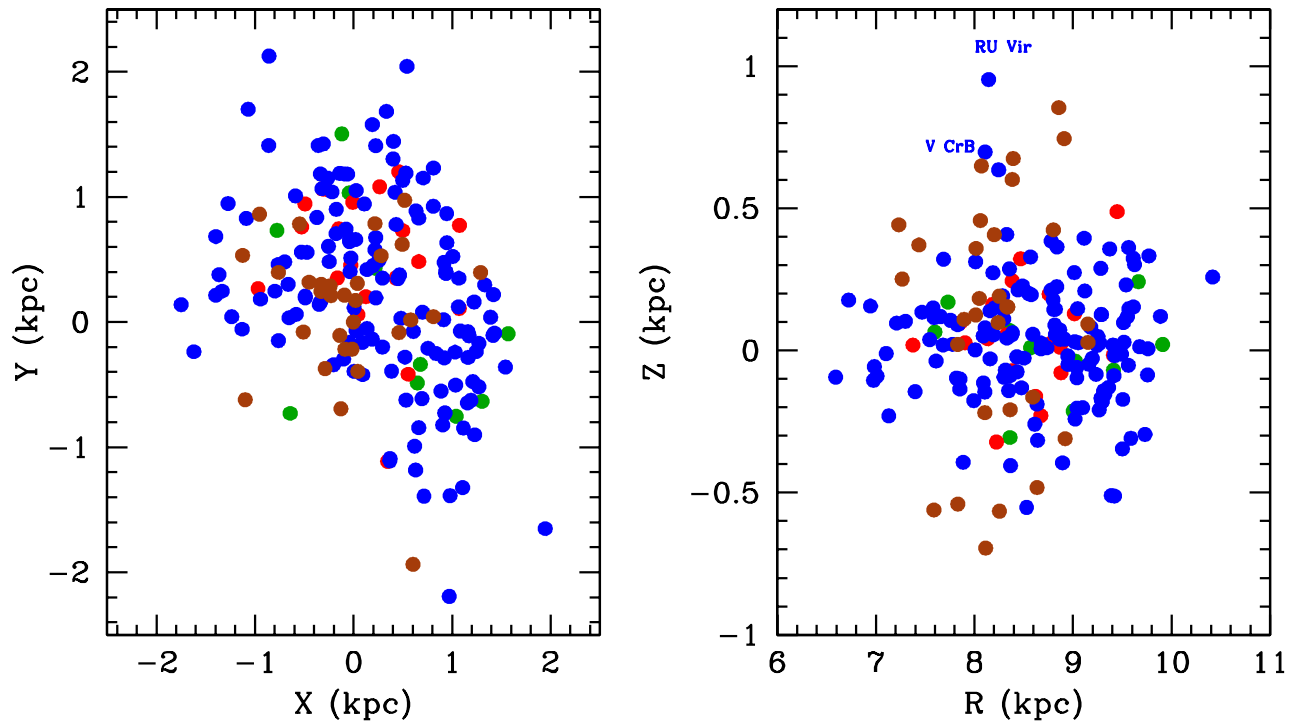}
   \caption{Left: Location of the sample carbon stars of different spectral types onto the Galactic plane. The Sun
   is placed at (X,Y)$=(0,0)$. Colour symbols: blue N-type stars; red J-type; green SC-type and brown R-hot type. Right: Distribution above/below the Galactic plane vs. the Galactocentric distance. The Sun is at R$=8.34$ kpc. The typical uncertainty in the (X, Y, Z) coordinates is $\pm 20$ pc, and $\pm 50$ pc for the Galactocentric distance. Two N-type stars with peculiar locations above the Galactic plane are labelled (see text).}
  \end{figure*}

\section{Gaia DR2 luminosities of carbon stars}
The determination of the bolometric correction (BC) for cool giants is still an open issue \citep[see e.g.][among many others]{bes84,bes98,mon98,cos96,hou00}. This is particularly challenging for AGB stars whose stellar flux is affected by millions of molecular absorptions. Furthermore, the formation of circumstellar shells around these evolved stars may absorb or scatter part of their stellar flux that is redistributed towards the mid and far-infrared ranges. Such complex processes still lack for an adequate description. It is therefore not surprising that in the extant literature the BCs adopted for  AGB stars may differ by up to 0.5 mag. In the present work, we have adopted the empirical BC$_K$ vs. $J-K$ relation for C-stars obtained by \citet{ker10}, which is based on a critical revision of the available studies. For $(J-K)$ between 1.0 to 4.4, the maximum standard deviation of this relation is 0.11 mag. We have used this BC relation also for the R-hot stars, which typically shows $(J-K)$ colours slightly bluer ($\sim 0.7-0.8$ mag) than the range of validity of the \citet{ker10} calibration. We note, however, that other BC$_K$ vs. $(J-K)$ calibrations found in the literature for $(J-K)<1$ values, provide BC$_K$ corrections differing by less than 0.1~mag \citep[e.g.][]{bes84,mon98} with respect to the one adopted here. Finally, for the O-rich AGB stars of the present study (see Table 1), we have also adopted the BC$_K$ vs. $(J-K)$ relation derived by \citet{ker10} for M stars. 

Then, we retrieved the 2MASS $J$ and $K_s$ photometry \citep{cut03} from the SIMBAD database and corrected them for the interstellar extinction according to the Galactic model of \citet{are92}, using the \citet{bai18}  derived  distances and the Gaia Galactic coordinates. For the reddening corrections, we used the relations $A_V=0.114 \cdot A_K$ and $A_J=2.47 \cdot A_K$. The corrected $J$ and $K_s$ magnitudes together with the calculated BC$_{K_s}$ values are given in Table 1 for all the sample stars. M$_{K_s}$ and M$_{\rm{bol}}$ magnitudes were then derived from the distance modulus relation and are also listed in Table 1. Uncertainties in M$_{K_s}$ and M$_{\rm{bol}}$ are dominated by those in $\varpi$ (distances).  For the typical parallax uncertainty in our stars ($\leq 10\%$, see Sect. 2) an error of $\sim\pm$0.20~mag in the absolute magnitudes is estimated.  Obviously, for the few stars in the sample  with parallax uncertainty within $10-20\%$, the error would  be larger. We therefore adopted $\pm 0.25$ mag as a typical error for M$_{K_s}$ and  M$_{\rm{bol}}$ in the full sample of stars. The corresponding bolometric luminosity distributions/functions (LF) obtained for all the carbon star types in our sample are shown in Figure 3.  Several remarks can be made:

a) The LF of the N-type stars peaks at M$\rm{_{bol}}\sim -5.2$ mag, with extended tails towards lower and higher luminosities. A Kolmogorov-Smirnov (KS) test says that the distribution is consistent with a normal one with a $p=0.6$. The average luminosity of our selected N-type stars is $\langle$M$_{\rm {bol}}\rangle=-5.09\pm 0.58$ mag. Within the quoted uncertainties, this average luminosity ($\sim10^4$~L$_\sun$) agrees with that found by similar studies of Galactic and LMC C-stars \citep[e.g.][]{gua13,gul12}  and with the theoretical expectations for stars with  a $\sim 0.6$ M$_\sun$ He-core mass \citep{pac71}. This occurrence confirms previous findings that the majority of the N-type carbon stars have an initial mass $2\pm0.5$ M$_\odot$. However, unlike previous  Galactic studies, we find a significant number of bright N-type stars  (M$_{\rm {bol}}\la -5.8$ mag), although none above the classical limit for AGB stars, M$_{\rm{bol}}=-7.1$ mag\footnote {Note that indeed luminous (M$_{\rm {bol}}\la -5.8$ mag) AGB carbon stars have been found in the Magellanic Clouds \citep{loo99}, some of them exceeding the classical AGB luminosity limit, which could be HBB stars in the latest stage of the AGB phase.}. In principle, this high luminosity tail of the LF should be populated by AGB stars with higher core masses and, hence, with higher initial mass (M$>3$ M$_\odot$). Note that stars with $3<M/M_\odot<5$ will be very luminous objects in the AGB phase and still become C-stars, but owing to the larger envelope dilution of the dredged up material, the time spent in the C-rich phase is quite short  thus, their contribution to the LF would be low. Moreover, in stars with mass M$>5$ M$_\odot$ the occurrence of the HBB prevents the formation of a C-rich envelope \citep[see, e.g.,][]{kar14} and the stars may exceed the AGB luminosity limit.  A more quantitative analysis of the  LF tails, both at low and high luminosity, will be illustrated in the next section. The derived average absolute $K$ magnitude is $\langle$M$_{K_s}\rangle=-8.16\pm 0.50$, which agrees well with values found in NIR photographic surveys of the Magellanic Clouds \citep[e.g.][]{fro80} and the Galaxy \citep{sch87}, although we find a larger dispersion. Nevertheless, this dispersion in M$_K$ is compatible with the typical range in T$_{\rm{eff}}$ (2500-3500 K) deduced for N-type stars. \citep[e.g.][]{berg01}.

\begin{figure*}
\centering
\includegraphics[width=17cm,angle=-00]{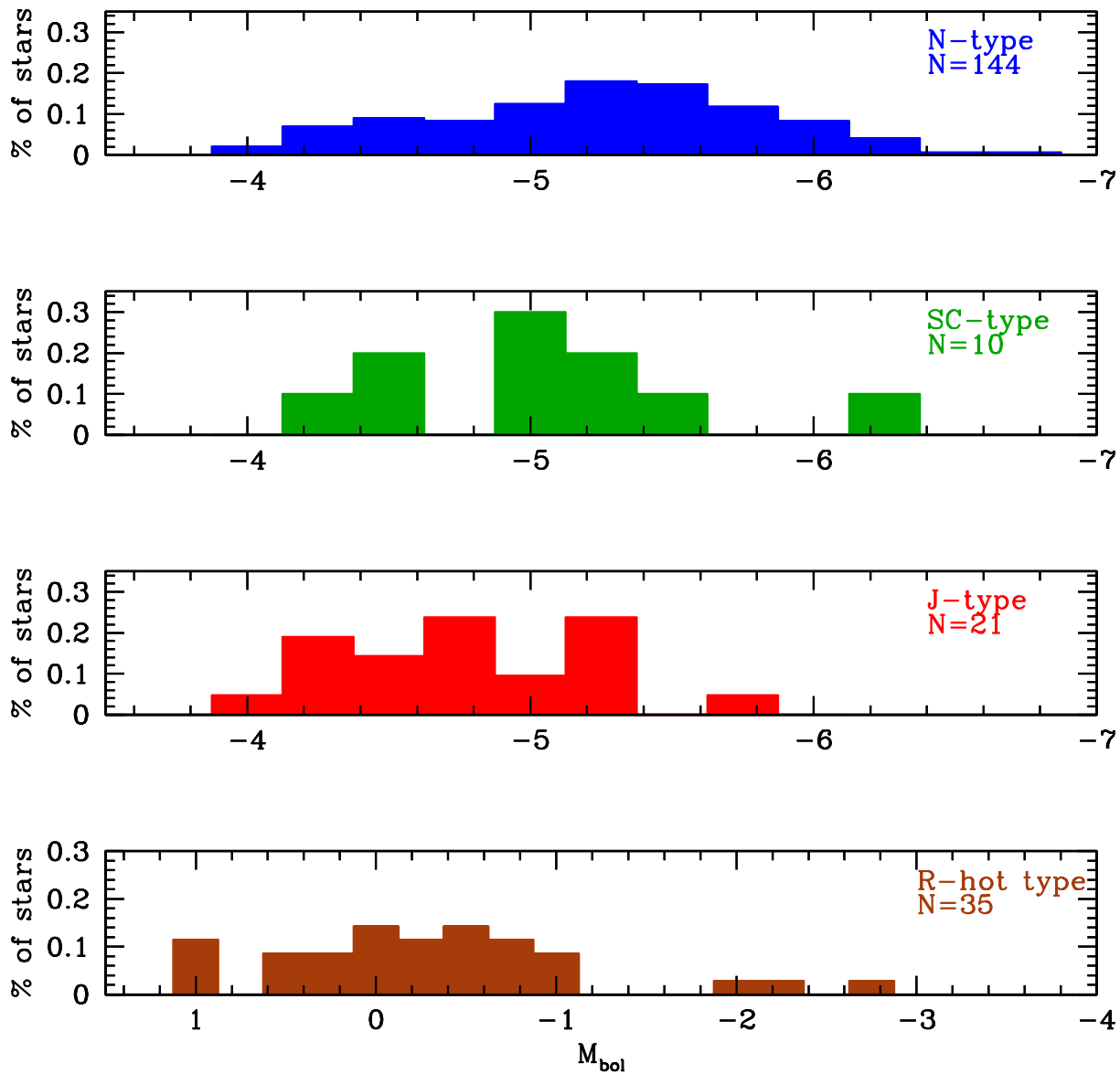}
\caption{Luminosity distributions derived in this study for the different spectral types of Galactic carbon stars. The bin-size is 0.25 mag. Note that the range in luminosity is different for the R-hot type stars that are fainter.}
\end{figure*}

b)  The LF of SC-type stars is rather similar to that of the N-type. This figure, together with the scarce number of SC-type carbon stars identified so far, is compatible with the hypothesis that the SC-type represent a short transition phase (C/O $\approx 1$) in between the O-rich (C/O$<1$) and the C-rich (C/O$>1$) AGB phases. 
%Note that intermediate
%mass stars ($>3$ M$_\sun$) that will never become carbon stars because of the operation of the HBB, theoretically may pass trough a very short C-rich phase due to the cessation of the HBB when the envelope mass is significantly reduced due to the strong mass-loss \citep[e.g.][]{kar14}. However, if that would be the case, the LF of SC-type stars should be biased toward higher luminosities with respect to N-type stars, which apparently is not seen in see Fig.~3. 
Moreover,  the very similar chemical composition shared by N- and SC-type carbon stars \citep{abi98,abi02,abi19}, as well as their similar location above/below the Galactic plane (see section 2), support the conclusion that both types of carbon stars originate from similar progenitors. In any case, the identification and analysis of more SC-type stars is needed to reach a definite conclusion. In the following, we will tentatively assume that SC- and N-type belong to the same stellar population. 

\begin{figure}
   \label{figobs}
   \centering
   \includegraphics[width=9cm]{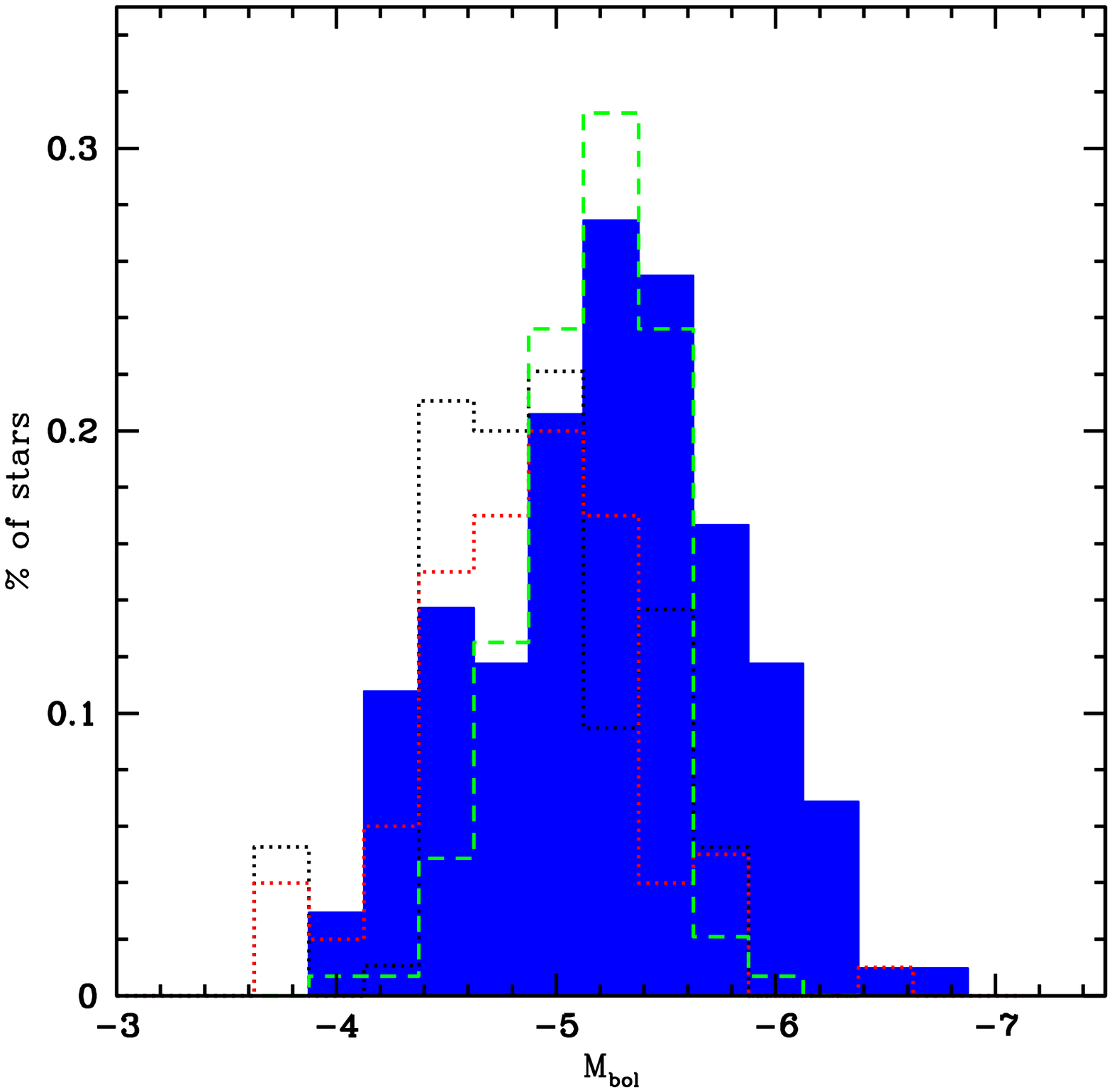}
   \caption{Derived LF distribution of the present study (blue histogram, excluding J and R-hot-type stars),
   compared to those obtained by \citet{gul12} (black dotted line) in the VMC survey of carbon stars,  \citet{gua13} (red dotted line) in Galactic AGB carbon stars, 
   and \cite{whi06} in Galactic Mira carbon stars (green dashed line). Note that the LF of these studies have been re-binned with a 0.25~mag step for a better comparison.}
  \end{figure}

c) J-type stars show a significantly dimmer LF compared to that of N-type carbon stars. Their average magnitudes are $\langle$M$_{\rm{bol}}\rangle=-4.66\pm 0.40$ mag and $\langle$M$_{K_s}\rangle=-7.70 \pm 0.40$ mag. Although these luminosites are within  the AGB range, a non-negligible fraction of the J-type  are fainter than M$\rm{_{bol}}=-4.5$ mag, which represents the threshold for the occurrence of TDU episodes \citep[see, e.g.][]{stra03b,cri11}.
Therefore, J-type stars not only differentiate chemically from N-type stars but also in their luminosity, which reinforces the idea that these stars have (as their carbon enhancement) a different origin. This is the first time that this conclusion is reached on the basis of an homogeneous luminosity study.  

d) As already noted, R-hot carbon stars show a LF much dimmer than that of the AGB stars: indeed we find  $\langle$M$_{\rm{bol}}\rangle=-0.10\pm 0.80$ mag and $\langle$M$_{K_s}\rangle=-2.54\pm 1.03$ mag, the latter being about half a magnitude brighter than that previously determined by \citet{kna01}. Note that the main reason of this discrepancy is the systematic difference between the Gaia DR2 and Hipparcos parallaxes (see Sect.~2).
This occurrence is at odds with their suggestion that these objects are He-burning red-clump stars. Indeed, the luminosity of the red clump is expected, for solar metallicity stars, at $\langle$M$_{K_s}\rangle\sim -1.6\pm 0.3$ mag \citep[e.g.][]{alv00,cas00,sal02}. On the contrary, our derived luminosities put the R-hot stars in the upper part of the RGB. Therefore, the hypothesis that the carbon enhancement in a significant fraction of the R-hot is produced by the C dredge-up powered by the violent off-centre He ignition (He flash) in the degenerate core of low-mass stars must be discarded. Furthermore, comparing the range of M$_{\rm{bol}}$ derived here for both the R-hot and the J-type stars, it seems also very unlikely that the latter type represents a luminous phase of R-hot stars, in the hypothesis that these are formed from a high-mass helium white dwarf subducted into a low-core-mass red giant  \citep{zha13}. The possibility that J-type stars are the  descendants of the R-hot ones in a more advanced evolutionary stage seems to be discarded.

In Fig.~4, we then compare the LF derived in the present study (N- and SC-type carbon stars) with the LF obtained for AGB carbon stars in three representative similar studies. We exclude the J-type and the R-hot stars since, as previously shown, these carbon stars probably have a different origin. In particular, we compare our derived LF with those of (i) the VISTA survey of AGB stars in the Large Magellanic Cloud (\citet{gul12}, 93 stars, black-dotted line)\footnote{We include here only those carbon stars in \citet{gul12} which they consider have dusty envelopes, and therefore are more probably placed on the AGB phase.} (ii) the sample of Galactic carbon stars by \citet{gua13} (102 stars, red-dotted line) and, (iii) the sample of Galactic Mira carbon stars by \citet{whi06} (145 stars, green-dashed line).
%We note that a KS test showed that is is unlikely that these three LF are consistent with a normal distribution, showing in the three cases a $p < 0.5$, at odd with the LF found here.

Globally, the comparison with  the \citet{gua13} and \citet{gul12} LFs reveals that they appear $\sim$0.25~mag dimmer as ours LF. In the case of the  \citet{gul12} LF and, since their BC$_K$ are very similar to ours, we ascribe this difference to the lower average metallicity of the carbon stars in the Large Magellanic Cloud. Indeed, at lower metallicity, owing to the lower O abundance in the envelope, the C-rich phase (C/O$>1$) is more rapidly attained ( i.e. at fainter luminosity) and lasts for a longer time.  
%Indeed, it is well know that the mass limit for carbon stars decreases with decreasing metallicity. At the typical  metallicities of [Fe/H]$\sim $-0.7 and $\sim$-0.4 (for the SMC and LMC, respectively), this lower limit might be as low as 1.2-1.3 M$_\sun$, compared to the $\ga$1.5~M$_\sun$ limit predicted at Solar metallicity \citep[e.g.][]{stra06,kar14}. This lower mass would shift the peak of the LF towards dimmer luminosities.
%OSCAR: The luminosity depends on the core mass. At low Z, even if the mass is lower the core mass is larger.
On the other hand, the differences with \citet{gua13} can be explained mainly by the shorter distances derived by these authors as compared to the ones we adopted: the average distance difference for the 21 stars in common is $210$ pc in the sense \citet{gua13} minus the present study,  with a signifcant dispersion ($\pm 240$ pc). Actually, these authors adopted  mainly Hipparcos parallaxes in their study. We remind that Hipparcos parallaxes tend to be larger than those of Gaia DR2 for carbon stars (see Section 2). Differences in the adopted BC probably play also a role: \citet{gua13} used a BC$_K$ vs. $K_s-[12.5]$ relation, where [12.5] is the stellar flux at 12.5 $\mu$m. In general we derived larger BC$_K$ than \citet{gua13} for the stars in common, which may be partially compensated by the extinction corrections not considered by these authors. 
This shift towards weaker M$_{\rm{bol}}$ is not seen in the LF derived by \citet{whi06} for Galactic carbon Miras (green dashed line in Fig. 4), which  peaks at the same M$_{\rm{bol}}\sim -5.2$ mag than ours. We recall that these authors derived their luminosities/distances on the basis of the period-luminosity relation found in the Miras of the LMC  corrected by the Galactic zero point \citep{fea06}. Their BC for the $K$-magnitudes were calculated from a relation with various infrared colours. In the particular case of the $(J-K_s)$ relation, BC may differ by up to 0.3~mag with respect to the relation adopted here for the reddest objects, i.e. for $(J-K_s)>2$ mag. For nine stars in common with these authors, we find a mean difference in the distance of  $120$ pc  in the sense \citet{whi06} minus this study, also with a significant dispersion ($\pm 250$ pc). In any case, from Fig. 4, it can be appreciated that their LF is very narrow around the luminosity peak, as opposed to  the other LFs shown in this figure. Moreover, their LF does not show any extended tails at low and high luminosity, unlike what we clearly find. This difference indicates that carbon Miras may represent a distinct stellar population among the AGB carbon stars. We will  come back to this issue in the next section.   

\subsection{Comparison with theoretical LF}

In Figure 5 we compare our new derived LF distribution with its theoretical counterpart. Details about the derivation
of such a theoretical distribution can be found in \cite{gua13} (see also \citealt{cri15}). Hereafter, we remind basic concepts only.

We extract the luminosities from our AGB models, sampling the C-rich phase with the same magnitude bins as the 
observational LF. We assign weights proportional to the time spent by the model in each magnitude bin. Then, we populate
our distribution by simulating a simple disk evolution with a Salpeter
initial mass function (IMF) and a metallicity distribution from \cite{cesco}. A linearly decreasing star formation rate has been considered. The contribution of each star to the LF is a function of the time spent in the C-rich phase. We have considered all stars attaining the C-rich regime (from 1.0 M$_\odot$ to 6.0 M$_\odot$, depending on the initial metallicity of the model).   It must be stressed that in the construction of the theoretical LF we consider the full evolutionary tracks of all the C-star models. These tracks already include, in particular, the post-flash luminosity dip, so that no correction is needed to account for this phenomenon \citep[see][]{ibe83,sta05}. 

The resulting theoretical LF
is shown in Figure 5 (red dashed histogram). Note that this new theoretical LF is shifted to slightly lower bolometric magnitudes with respect to those presented by \cite{gua13} (see Fig. 4, red dotted histogram). This is a consequence of the adoption of new bolometric corrections in the derivation of the mass-loss rate. The mass-loss prescription of FRUITY models are illustrated in \cite{stra06}. It consists of a fit, in the
mass-loss vs. period plane, to a sample of Galactic O-rich and C-rich giants. The stellar period is calculated by means of the M$_K$-period
relation proposed by \cite{white03} and M$_K$ is derived from the bolometric magnitude of the model and a BC$_K$ vs $T_{eff}$ relation. At variance with \cite{stra06}, here we have adopted an updated fit of the BC$_K$ vs T$_{eff}$ relation. In particular, we have considered the $T_{eff}$ data of O-rich red giants from 
\cite{buzzoni10}, extended at lower $T_{eff}$ with a sample of
N- and SC-type carbon stars from \citet{abi98} and \citet{abi02}. Then, the bolometric corrections have been obtained by means of the BC$_K$ vs $(J-K)$ relation of \citet{ker10} (see above).
The new and the previous relations are compared in Figure 6. Note that for $T_{eff}<3500$~K, which correspond to the typical temperatures of the C-stars, the new bolometric corrections are lower, and, hence,  the resulting M$_K$ is higher. This occurrence implies a reduction of the mass-loss rate and, in turn, an increase of the C-star lifetime. 
As shown in Fig.~7, in the case of the 2 M$_\odot$ model with Z$=$Z$_\odot$, the duration of the C-rich phase increases  ($+84$\%),  and the C/O ratio attains larger values 
($+18$\%). 
As a matter of fact, the new models experience a larger number of thermal pulses, thus spending more time at the higher luminosities. 

The peak of the present theoretical LF is in good agreement with the observed one. It is important to stress that intermediate mass models (M$> 4 $ M$_\odot$) show a larger sensitivity to the new mass-loss rate,
even if in our theoretical framework they rarely attain the C-rich stage (b.t.w. the O-rich TP-AGB phase of a 5 M$_\odot$ model with 
Z$=$Z$_\odot$ lasts 34\% longer and the C/O ratio increases by 55\%; see right panels of Figure 7). 

In spite of the good match of the central part of the LF, the theoretical prediction fails in reproducing both the low and the high luminosity tails of the observed LF. Interestingly, our theoretical LF agrees remarkably well  with the LF of Galactic Mira carbon stars (green dashed histogram in 
Figure 4).
We suspect that this feature is a direct consequence of the fact that our mass-loss rate is calibrated on a Mira M$_K$-period 
relation \citep{white03}. Different mass/loss prescription could be more appropriate for the irregular and semiregular pulsators which are the overwhelming majority in our sample, possibly making the brightest tail to appear. It this context, we noted that \citet{sta05} obtained a LF with an extended high luminosity tail by adopting the \cite{vw93} mass-loss prescription. According to these authors, this prescription causes significant mass loss only when the estimated pulsation period exceeds 500 days, while a negligible mass loss is obtained during the major part of the AGB evolution. As a result, the carbon star phase is more easily attained in the more massive models, those populating the high luminosity tails of the LF. In any case, the HBB should be inefficient in these stars, at least for those with mass M$\lesssim 5$ M$_\odot$, otherwise a C/O ratio well below unity  would be maintained within the envelope. In other words, the observed existence of C-stars with $-6.5<$ M$\rm{_{bol}}<-5.5$ severely constrains the occurrence of the HBB in intermediate-mass AGB stars\footnote{ Although models show that the HBB ceases when the envelope mass is reduced  down to $\sim 1$ M$_\odot$ \citep[see e.g.][]{kar14},  the duration of the C-star phase would be too short to provide any sizeable contribution to the LF, but very bright C-stars indeed exist (see e.g. IRAS04496-6958 in the LMC with M$_{\rm{bol}}\sim -6.8$, \citet{loo99,tra99}).}. 

Concerning the low-luminosity tail, it is likely due to the contamination of our sample with extrinsic C-stars, i.e. stars which inherited their  carbon enhancement from an already extinct carbon-rich AGB companion. Those extrinsic stars would approach the AGB
phase already with C/O$>1$ or very close to unity. Having lower core masses, they would show lower luminosities\footnote{We highlight that, at the beginning of the
TP-AGB phase of low mass solar metallicity models, the core mass is around 0.55 M$_\odot$, while they attain the C-rich phase when the core
mass is around 0.60 M$_\odot$.}.  This idea was already suggested by \citet{iza04} to explain the low luminosity tail of carbon stars in the SMC. On this same  basis, we run an exploratory case by accreting material with C/O$>1$ on a 0.7 M$_\odot$ main sequence star from a more massive ($3$ M$_\odot$) AGB
companion. After the accretion episode, the mass attained by the secondary star is 1.0 M$_\odot$. In Figure 5 we show the luminosity distribution of this model (dotted black curve), using as weight the time spent in each luminosity bin. As it can be seen, the low luminosity tail of the observed LF could be easily matched under the assumption that a fraction of these accreting stars become C-rich.

\begin{figure}
   \centering
  \label{figteo}
   \includegraphics[width=9cm]{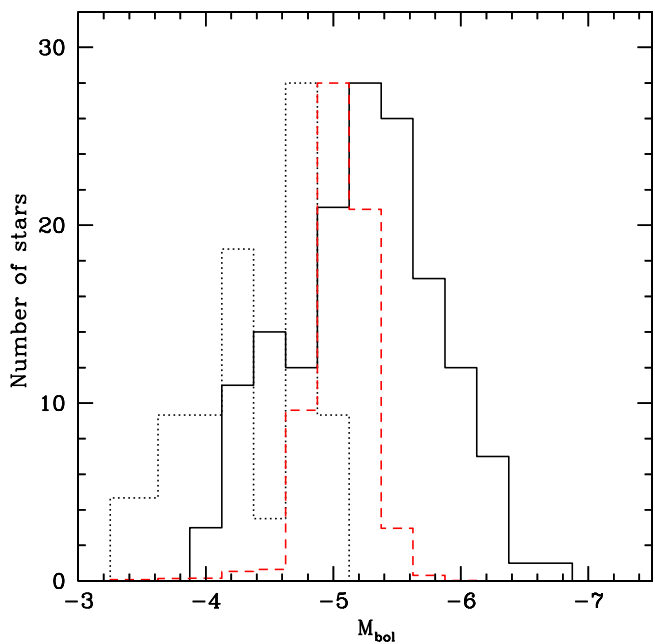}
   \caption{Derived LF for our full sample of Galactic AGB carbon (solid black line, excluding J-type and R-hot stars) compared with 
   the theoretical LFs described in Subsect.~3.1 (red dashed histogram), and to that LF expected for a typical model of an extrinsic C-star (black-dotted line, see text). The theoretical LFs are normalised such that the peak of a LF matches the peak of the corresponding observations.}
  \end{figure}

\begin{figure}
   \centering
  \label{newbck}
   \includegraphics[width=9cm]{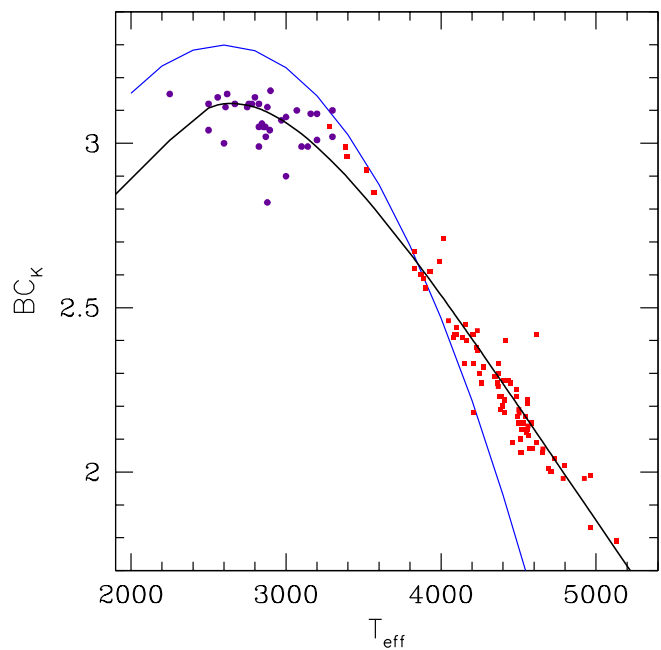}
   \caption{Bolometric corrections as a function of the effective temperature. Data are from \citealt{buzzoni10} (red squares) and this study (magenta circles). The new fit (black line) is compared with respect to the one adopted by \cite{stra06} (blue line). See text for details.}
  \end{figure}

 \begin{figure}
   \centering
  \label{moteo}
   \includegraphics[width=9cm]{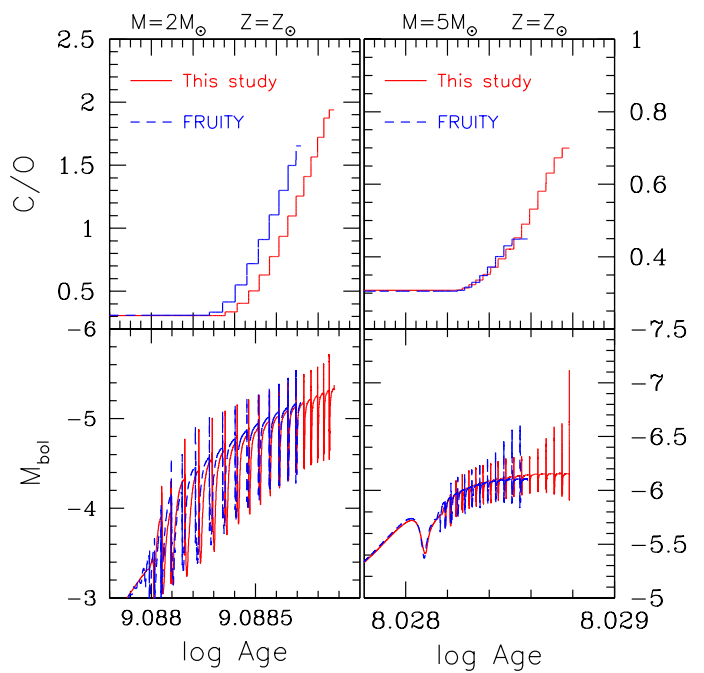}
   \caption{Evolution of the C/O ratio (upper panels) and surface luminosity (lower panels) for models calculated with different mass-loss laws. See text for details.}
  \end{figure}

%\begin{figure}
%  \label{fighr}
%%   \centering
%   \includegraphics[width=9cm]{Fig5bis.eps}
%   \caption{Location of the carbon stars (N- and %SC-types only) in the HR diagram, compared with FRUITY %tracks  of  the  corresponding mass and metallicity %(colour labelled).
%   The evolutionary tracks are limited to the %carbon-rich phase (C/O$>1$). Stars are plotted with %the same colour (black dots) for clarity. A typical %error bar is shown. See text for details.}
%   \end{figure}

\section{Kinematics}
Many  studies prior to Gaia have shown that the major stellar structures identified in our Galaxy - thin and thick discs and halo - have distinct kinematic and chemical properties at least in the Solar neighbourhood \citep[e.g.][]{Gilmore89,wys95,Freeman02,Bensby03, Kordopatis11,Ivezic12,Recio14,Wojno16}. Therefore, the accurate kinematic properties of our sample stars, thanks to the Gaia DR2 astrometric data, provide an additional and valuable piece of information to fully characterise the stellar population of the different types of carbon stars.  
 
 To compute the Galactocentric positions and velocities, we used the line-of-sight distance estimates of \cite{bai18}, together with the RA, DEC and proper motions of Gaia DR2 and our adopted line-of-sight velocities (V$_{\rm{rad}}$). Furthermore, we assumed the following parameters: (R$_\odot$, Z$_\odot) = (8.34, 0.025)$ kpc \citep{rei14}, (U$_\odot$, V$_\odot$, W$_\odot) = (11.1, 12.24, 7.25)$ km\,s$^{-1}$ \citep{sch12} and V$_{LSR}=240$ km\,s$^{-1}$ \citep{rei14}. The orbits were then computed using the \emph{galpy} code of \citet{bov15} assuming an axisymmetric potential using the Staeckel approximation of \citet{bin12}.
 
 The uncertainties on each of the derived parameters were estimated by adopting the standard
 deviation resulting from the propagation of 500 Monte-Carlo realisations on the line-of-sight distances, proper motions and radial velocities, hence from the consistent re-derivation of the positions, velocities and  orbits for each realisation. 
The final computed velocity components and the eccentricity of the orbits are reported in Table 1.

\begin{figure*}
   \centering
   \includegraphics[width=17cm]{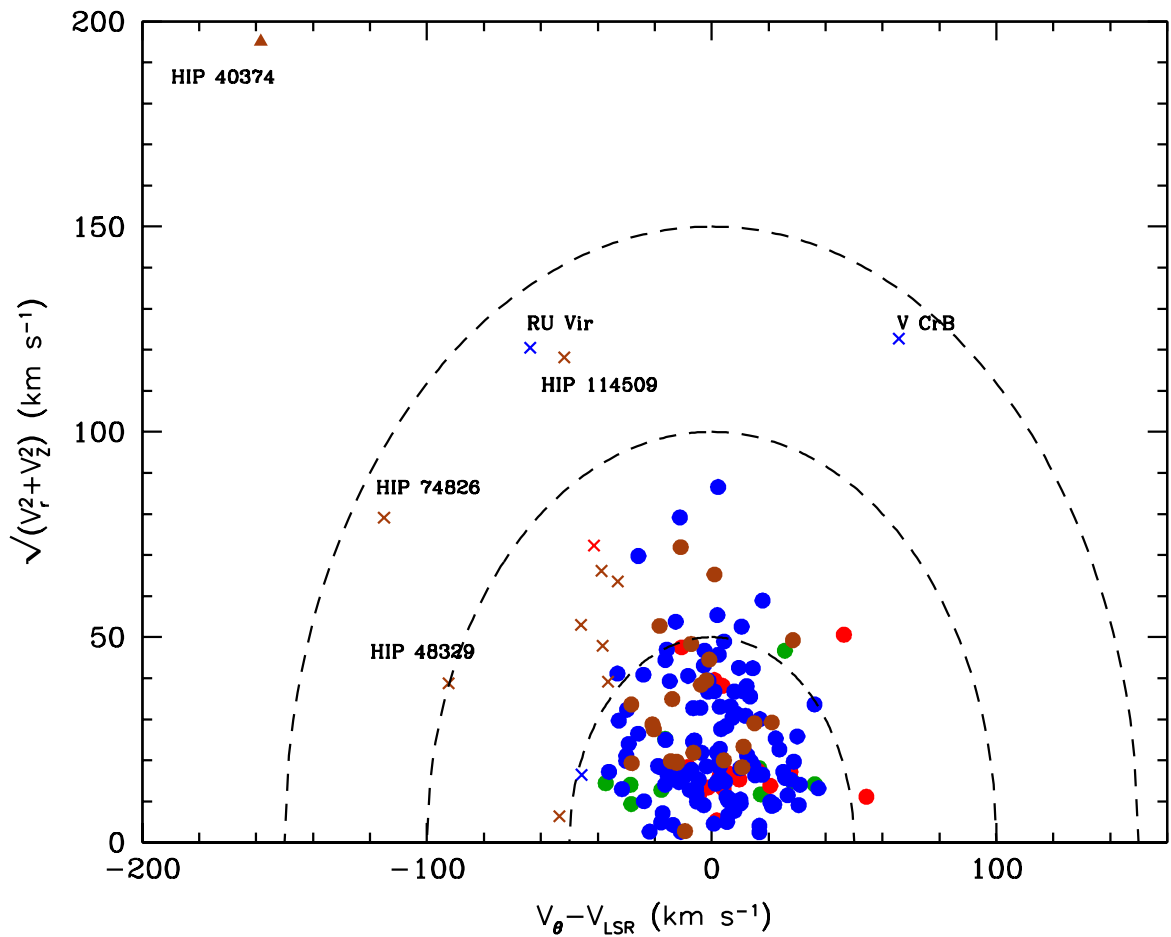}
   \caption{Toomre diagram for the selected stars of the present study. Colour symbols as in Fig. 2. Stars are plotted according to their membership probability ($>80\%$) to belong to the thin disk (solid circles), thick disk (crosses) or halo (triangles) stellar population. Dashed lines indicate $V_{\rm{tot}}^2= V_r^2+V_Z^2+(V_\theta-V_{\rm{LSR}})^2 = 50, 100$ and 150 km s$^{-1}$, respectively. Some stars with peculiar kinematics are labelled (see text).}
  \end{figure*}
  
  \begin{figure}
   \centering
   \includegraphics[width=9cm]{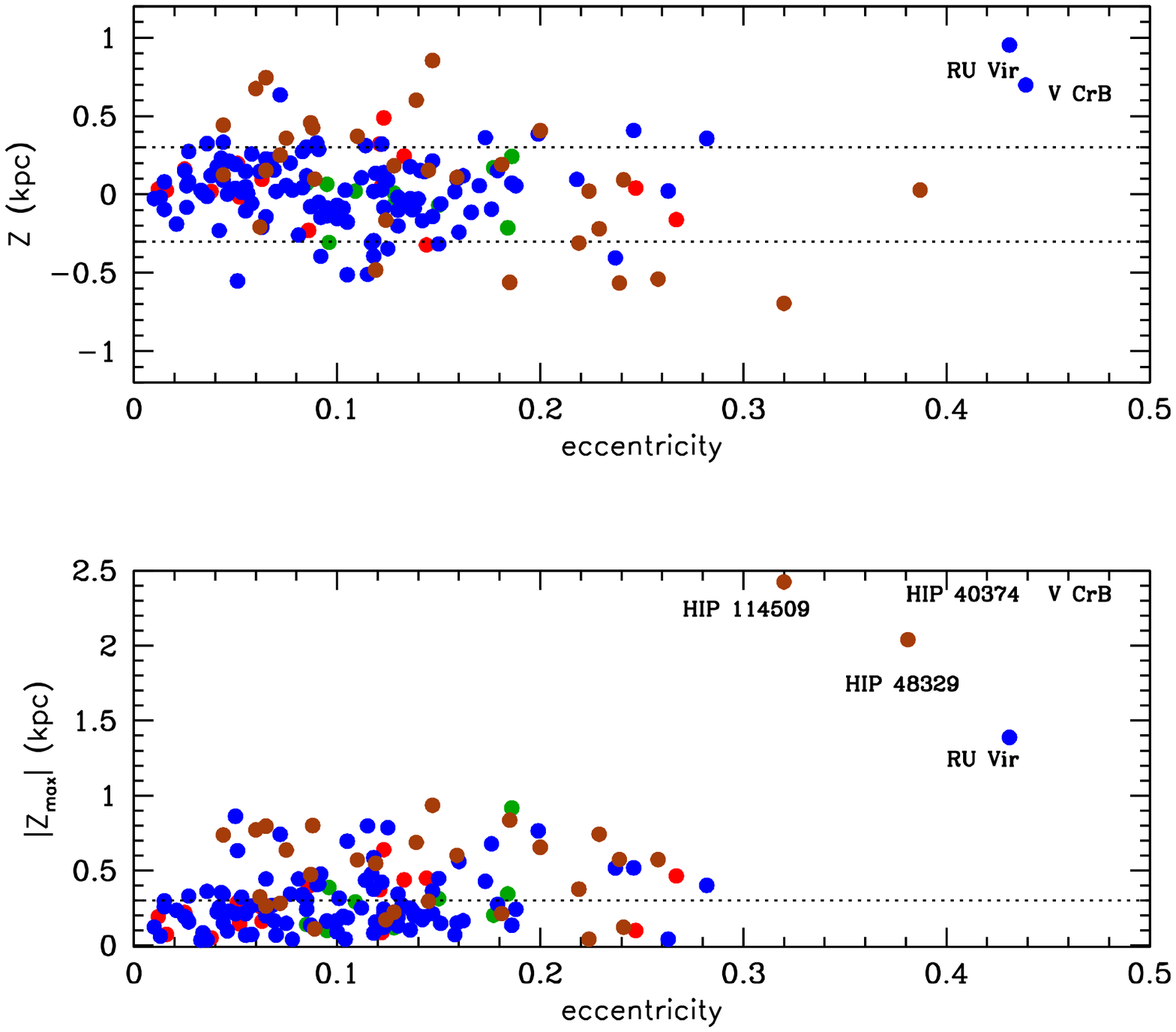}
   \caption{Top: Z-coordinate above and below the Galactic Plane vs. eccentricity. Bottom: estimated maximum Z coordinate along the orbit  vs. eccentricity. In the bottom panel the stars V CrB  and HIP 40374 are located beyond the Y-axis limits, at $|$Z$_{max}|= 4.4$ and 8.8 kpc, respectively. The horizontal dotted-lines in both panels at $Z\sim\pm0.3$ 
   indicate the commonly accepted scale-height of the local thin disc. Colour code as in Fig. 2.}
\end{figure}

A Toomre diagram, which is a representation of the stars combining their vertical and radial kinetic energies as a function of their rotational energy, is shown in Fig.~8. 
As a rule of  thumb, low-velocity stars (compared to the LSR) are likely to belong to the thin disc, high-velocity stars to the halo, and intermediate velocity to the thick disc. 
We computed for each star the likelihood, $L_i$, to belong to each of the Galactic components (where $i$ is associated to either the thin disc, the thick disc or the halo), using the following equation: 

\begin{equation}
\label{eq:proba}
L_i=\frac{1}{(2\pi)^{2/3}\sigma_{r}\sigma_{\theta}\sigma_{z}} \cdot \exp \left ( -\frac{V_r^2}{2\sigma_{r}^2} - \frac{(V_\theta - V_{i})^2}{2\sigma_{\theta}^2} - \frac{V_z^2}{2\sigma_{z}^2} \right )
\end{equation}

where $\sigma_{r}$, $\sigma_{\theta}$, $\sigma_{z}$  are the assumed velocity dispersion of each of the components (we omitted the $i$ subscript for visibility reasons) and $V_i$ is their mean rotational velocity (i.e. $V_\theta - V_{i}$ estimates the lag compared to the LSR). The assumed values are adopted from Table 6 in \citet{Kordopatis11}. We then compute

\begin{equation}
    P_i=\frac{L_i}{\Sigma_{i=1}^3 L_i}.
\end{equation}

If $P_i$ is greater than 80\% then we assign to the considered star the membership of the $i^{th}$ component that scored this probability. According to this criterion stars are quoted
with a 0, 1 and 2 in Table 1 depending if they belong to the thin, thick disc and halo, respectively. We find that N-, SC- and J-type carbon stars mostly belong to the thin disc, at a rate of 96\%, 100\% and 94\%, respectively. 
Two N-type stars (V CrB and RU Vir), both known as being Mira variables,  also have typical thick disc kinematics. Unlike RU Vir, where no metallicity information exists in the literature,  V CrB has two measurements: [Fe/H]$=-1.3$ derived by \citet{abi01} and [Fe/H]$=-2.12$ reported by  \citet{kip98}. Such a large difference clearly illustrates the problems of the spectroscopic analysis for such cool and chemically complex stars. That said, whether it is $-1.3$ or $-2.12$, such a low metallicity for a C-rich Mira is unusual for Galactic stars, and this could suggest that  V CrB is an extragalactic interloper \citep{fea06}. Indeed, there are carbon stars in the Galactic halo which are likely to have an extragalactic origin and, at least in the Sagittarius dwarf galaxy, some of them are known to be Miras \citep{iba01,whi99,Mauron14}. The additional  N-type star with probable thick disc membership (see Fig. 8 and Table 1) is V2309 Oph. However there is no information in the literature about its metallicity. \citet{bof93} found some Li enhancement in this star, log $\epsilon\rm{(Li)}\sim 1.0$. The sole J-type star belonging to the thick disc (BM Gem, see Table 1 and Fig. 8) according to our criterion has a metallicity [Fe/H]$\sim +0.2$ \citep{abi00}, which we note is typical of thin disc stars.

As far as R-hot stars are concerned, $\sim30\%$ of them have a probability higher than 80\% to belong to the thick disc, and one of them, HIP 40374, exhibits kinematics fully compatible with the halo  ($P_i=100\%$).  In fact, HIP 40374 is frequently considered in the literature as a halo CH-type star \citep{har85,aok97} showing a low $^{12}$C$/^{13}$C$\sim$10, a very common chemical property in CH-type stars. HIP 74826 is slightly metal-poor  ([Fe/H]$=-0.3$, \citet{zam09}), which is compatible with a thick disc membership.  At least six additional R-hot stars (some of them labelled in Fig. 8, see also Table 1) have kinematics compatible with the thick disc. Unfortunately no information exists in the literature about their chemical composition. Since the V$_z$ velocity component increases with the stellar age \citep[e.g.][]{nor04}, this figure reinforces our previous conclusion (see Section 2) that a significant part of the R-hot stars belong to an older (probably
less massive) stellar population than the other types of carbon stars.

\begin{figure}
   \centering
   \includegraphics[width=9cm]{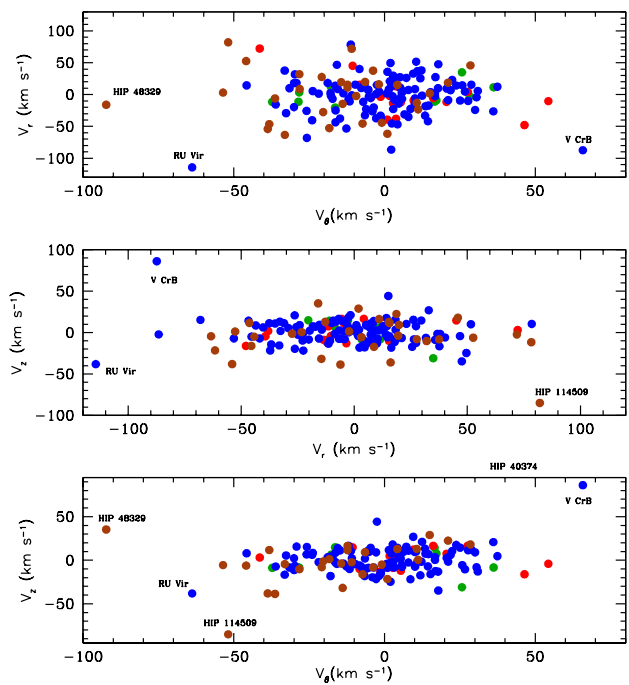}
   \caption{Velocity (radial, orbital and  vertical) diagrams for the sample
   stars. Orbital velocities are referred to the local standard of rest, V$_{\theta_{LSR}}=240$ km s$^{-1}$. Colour code as in Fig.~2. A few objects with peculiar kinematics are labelled.}
  \end{figure}

Figure 9 shows the location above/below the Galactic plane versus the eccentricity of the orbits for the different spectral types. Clearly, it is difficult to distinguish between the N-, SC- and J-types in this diagram. Most of  them have almost circular orbits ($e\lesssim 0.15$) and do not extend beyond $|z|\ga 300$ pc from the Galactic plane, this value being the typical scale-height of the thin disc \citep[e.g.][]{Juric08}. Interestingly enough, SC- stars have eccentricities only within 0.1-0.2. This result, if not related to small number statistics (we recall that we have only 10 SC-stars), is difficult to explain otherwise. Indeed,  considering that the vertical velocity of these stars is relatively low (see also Fig. 10), the fact that we do not find low eccentricity SC stars could imply that their orbits have been affected  by Lindblad resonances with the spiral arms \citep[e.g.][and references therein]{Sellwood14}.  However, in such a scenario, it would be puzzling why the other carbon stars in our sample have not been affected in a similar fashion. 
On the other hand, R-hot stars  deviate again from the above trend:   They  have on average a larger eccentricity ($\langle e \rangle, \sigma)= (0.17, 0.10$) than the other types of carbon stars; ($0.10,0.07),  (0.13,0.10), (0.10,0.08)$ for N-, SC- and J-types, respectively. Furthermore, 13 (3.6)\footnote{The number within the parenthesis indicates the possible deviation from this figure obtained as $\sqrt {N}$, where N is the number of stars.} of these R-hot stars out 35 are  located above $|z|\sim 500$ pc as opposed to  4 (2) stars out 144 for N-types, and zero out 10 and 21 for both SC- and J-types, respectively.  In fact, a similar fit to the $|z|$ distribution above the Galactic plane as already performed for N-type stars (see Sect. 2) gives  a scale height $z_o\approx 450$ pc. Some stars with extreme orbits are also indicated in Fig. 9 and 10. Note that, as expected, these stars are the same as those labelled in Fig. 8 (Table 1) with a possible thick disc membership.  Similar conclusions can be drawn from Fig. 10, which illustrates the relationship between the different velocity components: no clear distinction can be noticed between the different carbon star types with velocities typical of the thin disc stars, whereas the R-hot stars  show again more {\it heated } $V_r$ and $V_z$ distributions.

  \section{Characterisation of carbon stars with the Gaia-2MASS diagram}

\begin{figure*}
   \centering
   \includegraphics[width=17cm]{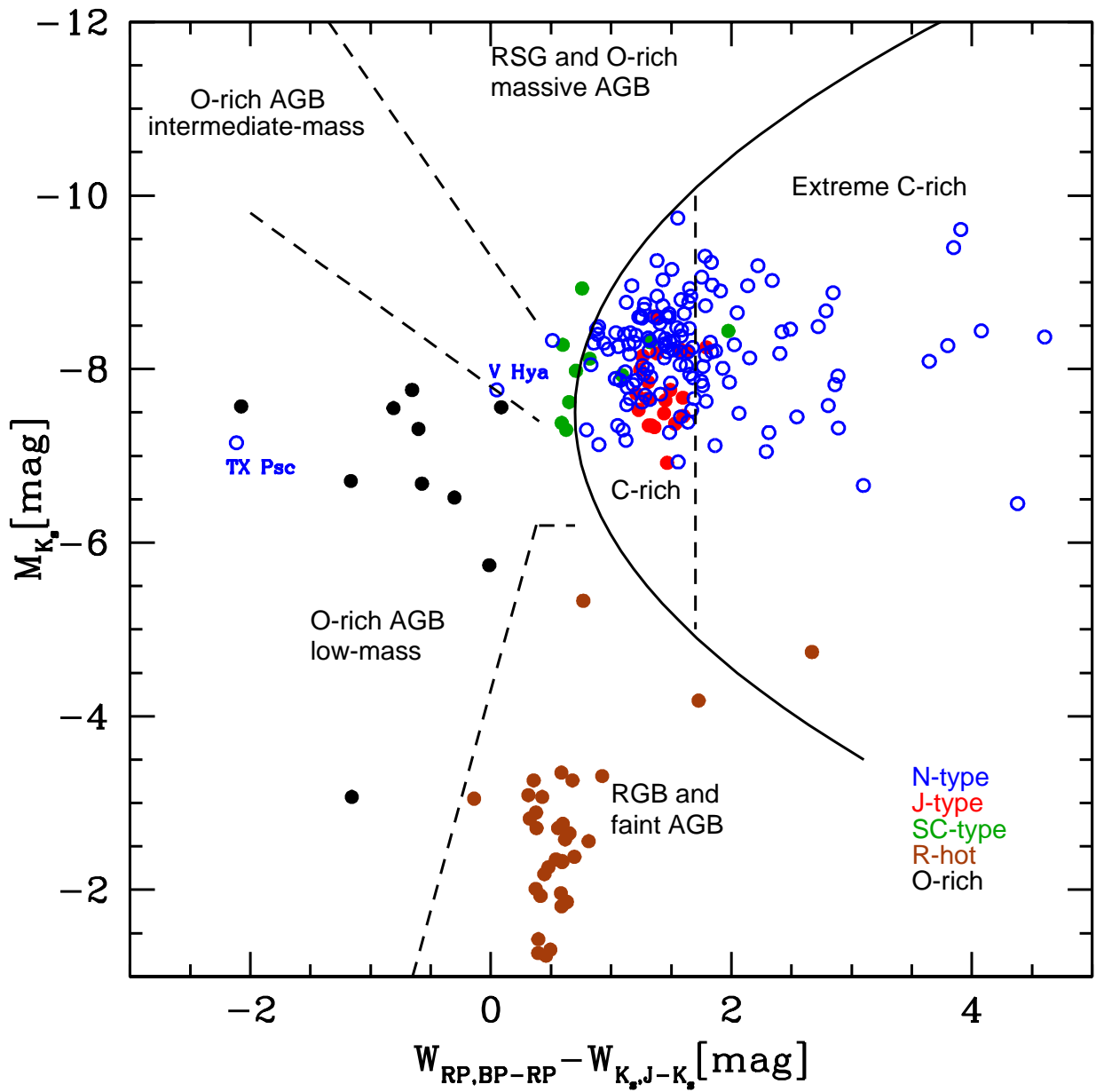}
   \caption{Gaia-2MASS diagram for our sample stars. The curved line delineates the theoretical limit between O-rich (left of the line) and C-rich AGB stars (right of the line). Dashed lines separate sub-groups of stars as indicated in the figure. Possible outlier stars are labelled (see text). We have used open blue circles for N-type stars for clarity. Colour code as in Fig. 2 (see text). The uncertainty in M$_{K_s}$ is $\pm 0.25$ mag typically. }
  \end{figure*}

 Recently, \citet{leb18}, using the Gaia DR2 LPV candidates in the Large Magellanic Cloud, showed that an optimal combination of visual and infrared photometry allows to  identify subgroups of AGB stars according to their mass and chemistry, i.e. C- and O-rich AGB stars. These authors adopted a particular combination of Gaia and IR photometry, leading to the infrared Wesenheit function defined as $W_{RP,BP-RP}-W_{K_s,J-K_s}$ with:
  \begin{equation}
  W_{RP,BP-RP}= G_{RP}-1.3(G_{BP}-G_{RP}),
  \end{equation}
 and 
 \begin{equation} 
  W_{K_s,J-K_s}= K_s -0.686 (J-K_s),
  \end{equation}
 where $G_{BP}$ and $G_{RP}$ are the Gaia magnitudes in the blue and the red band, respectively. In the so-called {\it Gaia-2MASS} diagram, which plots the $K_s$-absolute magnitude versus $W_{RP,BP-RP}-W_{K_s,J-K_s}$, the C- and O-rich AGB stars populate different regions, which allows to easily identify them as long as their distances are known. Furthermore, with the help of synthetic population models  \citep{gir05,mar17}, \citet{leb18} show that this diagram also allows the distinction between low-mass, intermediate-mass and massive O-rich AGB stars as well as supergiants and extreme C-rich AGB stars, the specific stellar mass range in each of these groups  depending on the stellar metallicity \citep[see][for details]{leb18}. The named extreme C-rich stars concerns very red objects with large $(J-K_s)$ and high mass-loss rates.  Recently, \citet{mow19} have extended this analysis to the Small Magellanic Cloud and the Galactic LPVs
 from the full Gaia DR2 archive, showing the power of this diagram to the study of populations harbouring AGB stars. In particular, they found some interesting features in the diagram emerging from the different metallicities between these three stellar systems.
 
 The distribution in the Gaia-2MASS diagram of our sample stars is shown in Fig. 11. For comparison purposes, we have added to the sample a few Galactic O-rich AGB stars (black filled circles,  see Sec. 2 and Table 1) presumably of intermediate mass ($\ga 3-4$ M$_\sun$) according to \citet{gar07}, fulfilling our parallax quality criterion (see  Section 2). Dashed lines delimit the regions of different stellar masses identified by
 \citet{leb18} (see their Table 1) for the LPV O-rich objects in the LMC, according to the synthetic population models by \citet{mar17}. We have adopted the same areas here but we have re-scaled them to the absolute magnitude $K_s$, assuming that the distance modulus of the LMC is 18.49 \citep{alv04}. Note that some differences may exists in these limits when taking into account the difference in metallicity between the main stellar population of the LMC ([Fe/H]$\sim -0.3$) and our Galactic stars ([Fe/H]$\sim 0.0$), but we do not expect them to  be significant. Nevertheless, \citet{mow19} found that such a difference is indeed observed at the limit separating O-rich and C-rich objects in the Gaia-2MASS diagram (continuous line in Fig. 11) depending on the average metallicity of the stellar population under study. This border is shown to be shifted to slightly larger $W_{RP,BP-RP}-W_{K_s,J-K_s}$ values with increasing average metallicity. According to these authors, this limit would be around  $W_{RP,BP-RP}-W_{K_s,J-K_s}\approx 0.9$ mag for Galactic LPVs, as it has been set in Fig. 11.  Moreover, due to the higher metallicity of the Galactic LPVs, the line delimiting C-rich and extremely C-rich objects ($W_{RP,BP-RP}-W_{K_s,J-K_s}\approx 1.7$ mag, in Fig. 11) should probably be slightly shifted to lower $W_{RP,BP-RP}-W_{K_s,J-K_s}$ values because the mass-loss rate is expected to be favoured for higher metallicities: in Galactic AGB carbon stars, larger mass-loss rates would appear earlier (lower $(J-K_s)$ value) during the AGB phase with respect to those in the MCs.
 
 Figure 11 also shows that the Gaia-2MASS can be used not only to separate O-rich from C-rich objects but also to identify the different types of carbon stars. N-type stars (blue circles) occupy clearly the whole C-rich region while the J-type stars (red dots) are located in the same region but shifted a bit to fainter M$_{K_s}$, as shown in Section 3. Remarkably, most of the SC-type stars (green circles), which are characterised by a C/O ratio very close to unity, are clearly located at the border between O-rich and C-rich objects. As far as the R-hot stars is concerned, they occupy a different region in the RGB and/or faint AGB area as compared to the other carbon stars spectral types\footnote{We checked that other carbon enriched objects (but not necessarily showing C/O$>1$ in their atmosphere), such as CH, Ba and CEMP stars are also located in this region.}. Moreover, a considerable number of N-type stars are located in the extreme C-rich zone. According to \citet{leb18}, these stars would have larger mass-loss rates and, preferably, larger C/O ratios. However, for the stars in our sample with derived mass-loss rate and C/O ratio in the literature, we have not found any correlation between these and the increasing $W_{RP,BP-RP}-W_{K_s,J-K_s}$ value. Nevertheless, the N-type stars CW Leo and IY Hya (initially in our sample but excluded because of their  parallax uncertainty), which are known to be extremely red objects with huge mass-loss rates ($\ga 10^{-5}$ M$_\sun$yr$^{-1}$), are indeed located in this region with $W_{RP,BP-RP}-W_{K_s,J-K_s}>4$ mag. Clearly, larger and more complete stellar samples are needed to draw a firm conclusion.
 
 The N-type stars TX Psc and V Hya (see Fig. 11) are found clearly outside their expected location. Both stars are relatively well-studied carbon stars. The average C/O ratio derived in TX Psc is 1.07 \citep{klo13}, which has been interpreted as evidence of a recent transition from an oxygen-rich atmosphere to a carbon-rich, making TX Psc a relatively “fresh” carbon star. \citet{bru19} have also recently detected an elliptical detached molecular shell around this star and speculate about the possibility of a recent (or a few) TDU episode that transformed it into a C-rich object. Thus, it could be that TX Psc became recently a carbon star and is actually  moving rapidly toward the right of the Gaia-2MASS diagram;  we have just "caught" it  in this move\footnote{We note that TX Psc is suspected to be a binary star and that it is the brightest N-type star in our sample ($G=3.72$ mag). Such a bright G-magnitude may be affected by saturation problems that could occur for the brightest Gaia DR2 targets, also affecting the determination of the parallax \citep{dri19}.
 These two facts may, therefore, affect its  
 $W_{RP,BP-RP}-W_{K_s,J-K_s}$ value.}.  V Hya shows evidence for high-velocity, collimated outflows and dense equatorial structures. It is considered as a transition object from an AGB star toward an aspherical planetary nebula \citep{sci19}. The observed circumstellar structures are also compatible with the existence of a binary companion in an eccentric orbit \citep{sah16}. The impact of the presence of circumstellar structures and binarity on the location of carbon stars in the Gaia-2MASS is out of the scope of the present study, but the cases of TX Psc and V Hya are quite promising for the use of this diagram also as a tool to identify "peculiar" AGB stars. 
 
 Finally, the few O-rich AGB stars in Fig. 11 are located where they were expected to be. Nevertheless, several studies \citep{gar06,gar07,per17} point out that some of these stars are intermediate rather than low-mass O-rich AGB stars, i.e. they should be found in Fig. 11 at larger $W_{RP,BP-RP}-W_{K_s,J-K_s}$ values and brighter $K$-luminosity. Indeed, they show strong Li lines and enhanced Rb abundances, which is expected to be found in intermediate-mass ($\ga 3-4$ M$_\sun$) AGB stars experiencing HBB. Obviously, the scarce number of O-rich objects in Fig. 11  prevents us to extract any definitive conclusion. 
 
 %In summary, we have shown the {\bf powerful} utility %of the Gaia-2MASS diagram to identify the different %types of carbon stars. Our first study should be %followed and completed by more detailed spectroscopic %analyses.

\section{Summary}
In this study, we have analysed the Gaia DR2 data for a sample of
210 luminous red giant carbon stars of N-, SC-, J- and R-hot spectral types belonging to the Solar neighbourhood  and fulfilling the criterion {\bf $\epsilon(\varpi)/\varpi\leq 0.2$}, in order to derive accurate luminosities and kinematic properties. The sample contains variable stars of Mira, irregular and semi-regular types. Our conclusions can be potentially affected by the accuracy of the distance estimation from Gaia DR2 data and the limited statistics in the case of the SC- and J-type stars. They can be summarised as follows:

a) We have shown that carbon stars of types N, SC and J are distributed homogeneously and in a similar way within $\sim 1.5$ kpc from the Sun. A simple exponential fit to the distance from the Galactic plane results in a scale-height $z_o\approx 180$~pc, in agreement with previous determinations. This scale height is compatible with a mass 1.5-1.8 M$_\odot$ for the progenitors of AGB carbon stars at solar metallicity, also in agreement with theoretical expectations.  In contrast, we confirm that R-hot stars extend more above/below the Galactic plane, which indicates a lower mass and older population for them.  

b) N- and SC-type carbon stars are well represented by a population having
a Gaussian distribution of absolute magnitudes such that (M$\rm{_{bol}},\sigma_o) = (-5.09, 0.58)$. Our derived bolometric luminosity function shows two tails at fainter and brighter luminosities more extended than previously found. This luminosity function contrasts with the much narrower one derived in Galactic Mira carbon stars \citep{whi06}, which would constrain this variable type of carbon stars to a mass range around $\sim 2$ M$_\odot$. The derived LF implies that  Galactic AGB carbon stars may reach up to M$\rm_{bol}\sim -6.0$ mag, similar to that found in the Magellanic Clouds. We show that the fainter tail can be understood by considering a contribution of extrinsic low mass (M$\la 1$ M$_\odot$) carbon stars stars, while the brighter tail would imply that a non negligible fraction of stars with initial masses up to 5 M$_\odot$ may become carbon stars during the AGB phase. This would increase the lower mass limit for the occurrence of the HBB to this value, which is in some tension with the extant theoretical models. 

c) J-type stars are clearly fainter than the N- and SC-types by about half a magnitude both in M$_{\rm{bol}}$ and M$_K$. This is the first time that this figure is clearly shown on an observational basis. This result, together with their very different chemical pattern, confirms that these objects have a different origin, otherwise unknown yet.   

d) We find that R-hot stars have an average $\langle$M$_K\rangle=-2.54\pm 1.53$ mag, which places these stars in the upper part of the red giant branch. This is about half a magnitude brighter than the value found by \citet{kna01} on the basis of Hipparcos parallaxes. This makes it unlikely for these stars to be red-clump objects and the progenitors of J-type stars as previously suggested.
%and, therefore, would discard a possible anomalous He-flash as the origin of %their carbon enhancement.  Furthermore, comparing the range of M$\rm{_{bol}}$ %derived here for both the R-hot and the J-type stars, it seems also rather %unlikely that the latter type represents a luminous phase of R-hot stars, in %the hypothesis that these are formed from a high-mass helium white dwarf %subducted into a low-core-mass red giant as recently suggested by %\citet{zha13}.

e) Our kinematic study shows that the overwhelming majority of the N-, SC- and J-type stars belong to the thin disc population. At contrary, a significant fraction ($\sim 30\%$) of the R-hot stars have kinematics typical of the thick disc, which points towards lower masses and older ages for these objects with respect to the other carbon star types.  

 Finally, we have shown the potential of the Gaia-2MASS diagram recently used for the LPVs in the Large Magellanic Cloud, to identify the different types of carbon stars, which opens an amazing opportunity to identify them in any stellar system as far as accurate distances are known.  

\begin{acknowledgements}
  This work has been partially supported by the Spanish grants AYA2015-63588-P and PGC2018-095317-B-C21 within
  the European Funds for Regional Development (FEDER) and by the "Programme National de Physique Stellaire"
  (PNPS) of CNRS/INSU co-funded by CEA and CNES. It has made use of data from the European Space Agency (ESA)
  mission Gaia (https://www.cosmos.esa.int/gaia), processed by the Gaia Data Processing and Analysis Consortium
  (DPAC, https://www.cosmos.esa.int/web/gaia/dpac/consortium). Funding for the DPAC has been provided by national
  institutions, in particular the institutions participating in the Gaia Multilateral Agreement. This research
  has also made use of the SIMBAD database, operated at CDS, Strasbourg, France.  The authors would like to thank
  the referee, J.T. van Loon, for the careful reading of the manuscript and useful 
comments and suggestions.

\end{acknowledgements}

\longtab{
\begin{landscape}
\begin{longtable}{lccccccccccccccc}
\caption{Derived data for the sample stars}\\
\hline
\hline
Star & GDR2 source id & $J_o$ & $K_{s_o}$ & BC$_{K_{s_o}}$ &  d  & M$_{\rm {bol}}$ & V$_{\rm {rad}}$  &  R    & Z    &  V$_r$       & V$_\theta$     & V$_z$       & e & P & Ref. \\
     &                &       &          &             & (pc)&               & (km s$^{-1})$  & (kpc) & (kpc)& (km s$^{-1})$ & (km s$^{-1})$ & (km s$^{-1})$&   &   &\\
\hline
\endfirsthead
\caption{continued.}\\
\hline\hline
Star & GDR2 source id & $J_o$  & $K_{s_o}$ & BC$_{K_{s_o}}$ &  d & M$_{\rm {bol}}$ & V$_{\rm rad}$  &  R & Z  &  V$_r$ & V$_\theta$  & V$_z$  & e & P & Ref. \\
& & & & & (pc) & & (km s$^{-1})$& (kpc) & (kpc)& (km s$^{-1}$)& (km s$^{-1}$) & km s$^{-1}$&&&\\
\hline
\endhead
\hline
\endfoot
SC-type & & & &&&&&&&&&& &&\\
WZ Cas  &429338816550169 &1.88&0.49&2.99&482.6&$-$4.83&$-$34.0&8.56&0.01&3.4&211.7&$-$8.7&0.12& 0 &    1 \\
CY Cyg  &216635570425088
 &3.00&1.96&2.78&1508.9&$-$6.06&2.7&8.35&0.07&$-$8.9&257.2&7.6&0.08& 0 &     3\\
RZ Peg  &190004711604318
 &4.00&2.79&2.89&1084.4&$-$4.39&$-$32.1&8.36&$-$0.30&$-$20.4&223.6&14.9&0.10& 0 &    1\\ 
FU Mon  &312507222237735
 &2.91&1.42&3.04&761.0&$-$4.84&26.2&9.02&$-$0.03&$-$11.3&211.5&$-$8.3&0.13& 0 &    1\\ 
AM Cen  &606589484459292
 &3.05&1.68&2.99&983.1&$-$5.19&22.0&7.73&0.17&$-$11.5&202.8&$-$8.7&0.18& 0 &    6 \\
R CMi  &315931527507378
 &3.61&2.39&2.89&1465.6&$-$5.45&46.1&9.66&0.24&34.9&265.7&$-$31.0&0.19& 0& 1\\
BB Tau  &342912189079226
 &4.46&2.86&3.09&1573.0&$-$4.93&2.1&9.91&0.02&$-$11.1&256.6&14.3&0.11&   0&   3\\ 
V372 Mon  &31038072895409
 &4.60&2.93&3.12&1288.0&$-$4.40&14.5&9.41&$-$0.06&$-$2.8&268.3&14.6&0.15&   0&   3 \\
V776 Mon  &299916137344349
 &3.96&2.34&3.10&844.8&$-$4.10&19.9&9.00&$-$0.21&11.4&276.2&$-$8.3&0.18&   0&   3 \\
IRC +10397&431098904018602
 &3.49&1.83&3.11&1068.1&$-$5.10&$-$21.0&7.59&0.06&11.0&222.3&6.4&0.09&   0&   3  \\
&&&&&&&&&&&&&&\\
J-type & &&&&&&&&&&&&&\\
BM Gem  &870169071480226
 &4.16&2.70&3.03&1109.1&$-$4.50&93.0&9.37&0.35&72.3&198.6&3.0&0.28&  1 &   5\\ 
WX Cyg  &206065281388967
 &3.51&2.05&3.03&1095.7&$-$5.12&46.3&8.13&0.04&$-$10.3&294.4&$-$4.1&0.25& 0  &  2\\ 
RY Dra  &167884430874641
 &1.55&0.19&2.98&381.3&$-$4.74&$-$20.0&8.46&0.32&$-$38.1&243.8&1.9&0.12&   0&   5\\ 
V Cyg  &216759128043799
 &2.92&0.05&3.10&456.3&$-$5.15&$-$12.0&8.32&0.05&$-$39.3&241.0&$-$4.5&0.12&  0&   2 \\
VX And  &385742630743259
 &2.77&0.94&3.16&612.2&$-$4.83&9.8&8.62&$-0$.16&$-$47.9&286.4&$-$16.1&0.27&   0&   1\\ 
Y Cvn  &154255362337068
 &0.74&$-0$.81&3.07&231.4&$-$4.56&15.3&8.38&0.24&4.2&267.6&16.6&0.13&0 &1\\
S Cam  &48478670689396
 &4.14&2.51&3.10&1403.9&$-$5.12&$-$13.0&9.44&0.48&$-$11.6&260.6&7.5&0.12& 0 &    5\\ 
V353 Cas &199721521490011
 &4.67&2.79&3.18&1288.9&$-$4.59&--   &8.87&$-0$.07&--&--&--&--& --&\\
UV Cam  &474001975295012
 &2.93&1.92&2.76&826.3&$-$4.91&$-$10.5&9.01&0.12&$-$9.1&249.7&12.3&0.06&0  &    1\\ 
FO Ser  &409810417145455
 &3.84&2.66&2.88&1004.8&$-$4.48&$-$16.2&7.37&0.01&12.4&244.1&4.1&0.04& 0  &   1 \\
HO Cas  &522753393163282
 &4.43&2.81&3.10&884.3&$-$3.82&--&8.86&0.01&--&--&--&--&-- &\\
V623 Cas &460202348450738
 &2.56&1.08&3.04&556.2&$-$4.61&$-$7.4&8.77&0.02&$-$2.1&241.7&4.9&0.02&0  &    1\\ 
V614 Mon  &310863417994966
 &3.11&1.72&3.00&694.1&$-$4.49&22.9&8.90&0.03&$-$3.5&238.7&$-$12.9&0.01& 0 &    1 \\
T  Lyr  &209618593730528
 &2.19&0.23&3.19&412.4&$-$4.65&$-$12.0&8.18&0.16&$-$4.5&235.6&10.9&0.03&  0 &   1 \\
UZ Pyx  &564214124217766
 &3.20&1.74&3.03&1179.0&$-$5.58&13.0&8.75&0.19&$-$12.0&245.4&$-$12.0&0.05& 0  &   1 \\
CG Vul  &452010752748216
 &3.28&1.62&3.11&928.0&$-$5.10&$-$20.0&7.85&0.09&15.3&231.8&$-$10.1&0.06&  0 &   4\\
IRC +30392 &20270020543655
 &4.70&2.77&3.19&1065.6&$-$4.18&--&7.90&0.02&--&--&--&--&--&\\
V429 Cyg &20589783264147
 &4.48&2.56&3.19&1110.5&$-$4.48&--&8.09&0.05&--&--&--&--& --&\\
V624 Cyg  &196791350920567
 &4.66&2.45&3.22&960.8&$-$4.24&--&8.38&$-0$.07&--&--&--&--&--& \\
RX Peg  &17947967434284
 &4.03&2.28&3.14&835.7&$-$4.19&$-$27.0&8.22&$-0$.32&45.1&229.5&14.7&0.14& 0  &    4\\ 
VY And  &193607144629589
 &4.64&2.94&3.13&1142.0&$-$4.22&$-$7.0&8.67&$-0$.22&$-$5.5&256.2&16.5&0.09&  0 &   1 \\
UV Aur  &180919213811384
 &3.67&1.98&3.12&1078.0&$-$5.06&$-$5.6&9.41&$-0$.01&$-$13.6&242.8&$-$8.8&0.05&  0&    4 \\
&&&&&&&&&&&&&&&\\
N-type &&&&&&&&&&&&&& &\\
R Scl &501613814518625
 &1.41&$-0$.03&3.02&436.1&$-$5.21&$-$5.4&8.36&$-0$.40&$-$68.1&214.2&15.3&0.24&  0 &   1\\ 
AQ Sgr  &418040440779451
 &2.39&0.73&3.11&556.3&$-$4.88&9.7&7.85&$-0$.13&$-$8.7&261.1&$-$1.7&0.10&  0 &   1 \\
TX Psc  &27430041294292
 &0.82&$-0$.62&3.02&201.4&$-$4.13&11.5&8.34&$-0$.14&$-$41.8&254.4&$-$7.1&0.15&0 &1\\
R Lep &298708272281571
 &2.08&0.07&3.20&414.0&$-$4.81&16.7&8.63&$-0$.18&$-$3.2&234.8&9.4&0.02& 0 &   2 \\
RT Cap  &685396678013396
 &2.20&0.53&3.12&426.1&$-$4.50&$-$29.2&7.99&$-0$.17&18.3&221.2&3.4&0.10& 0  &   1\\ 
RV Cyg  &195283085535929
 &1.74&0.33&3.00&522.5&$-$5.26&2.0&8.32&$-0$.07&$-$17.2&254.6&7.0&0.09&   0&   1  \\
S Sct  &425199357123128
 &1.66&0.19&3.04&370.9&$-$4.62&$-0$.2&8.00&0.00&$-$5.9&250.2&$-$7.4&0.05&    0&  1\\ 
SS Vir  &369985470761623
 &2.31&0.58&3.14&685.0&$-$5.46&2.0&8.24&0.63&$-$15.4&250.0&9.4&0.07&0    &  4 \\
ST Cam  &483958671558729
 &1.77&0.24&3.06&590.0&$-$5.55&$-$12.0&8.79&0.17&$-$6.7&245.8&$-$7.6&0.04& 0   &  1\\ 
TU Gem  & 3427006189904854656&1.85&0.62&2.91&1178.8&$-$6.83&48.0&9.51&0.09&31.7&210.3&$-$6.4&0.16& 0  &   4 \\
TW Oph  &412169598448745
 &2.27&0.33&3.19&590.6&$-$5.34&2.7&7.75&0.10&$-$9.3&265.1&14.4&0.11&  0 &   1\\ 
U Cam  &486859664266293
 &1.99&0.31&3.12&602.7&$-$5.47&4.4&8.81&0.08&12.6&242.2&7.9&0.04&  0 &   1 \\
UU Aur  &944939847899351
 &0.99&$-0$.47&3.03&497.1&$-$5.92&13.4&8.82&0.14&10.0&208.5&$-$8.3&0.14& 0  &   1 \\
UX Dra  &228951038675395
 &2.00&0.20&3.16&485.0&$-$5.07&4.8&8.48&0.22&$-$1.7&252.5&21.1&0.07& 0    &  1 \\
V460 Cyg  &195079407879237
 &1.75&0.23&3.06&415.8&$-$4.81&8.7&8.31&$-0$.06&9.5&260.7&2.6&0.10&   0 &  1\\ 
VY UMa  &10600057669689
 &1.95&0.45&3.05&425.9&$-$4.65&$-$6.3&8.56&0.32&$-$18.3&253.6&7.5&0.09& 0&1\\
Y Hya &566385292032574
 &2.02&0.50&3.06&470.0&$-$4.80&3.5&8.44&0.21&$-$46.8&244.3&14.3&0.15& 0  &   1\\ 
Y Tau  &340001774928534
 &1.80&0.17&3.10&612.8&$-$5.67&17.0&8.94&$-0$.01&1.8&240.7&$-$4.2&0.01&  0 &   1 \\
Z Psc  &29473409480479
 &2.24&0.82&3.02&571.0&$-$4.95&18.0&8.64&$-0$.31&$-$4.3&270.6&$-$8.0&0.15&  0 &   1\\ 
W CMa  &30453738815229
 &2.01&0.85&2.82&550.9&$-$5.07&23.0&8.73&0.00&3.1&245.7&$-$5.8&0.03& 0 &1\\
SZ Sgr  &412015543131265
 &3.44&2.03&3.01&1420.3&$-$5.72&16.5&6.94&0.15&$-$22.8&243.0&$-$1.4&0.07& 0  &   5 \\
RX Sct  &425280869160341
 &2.83&1.40&3.02 &728.3&$-$4.89&$-$8.0 &7.68 &0.01&14.6 &265.8 &$-$5.7&0.12 & 0  &   1 \\
V CrB  &137693387764271
 & 3.22&1.45&3.15&823.6&$-$4.98&115.0&8.10&0.69&$-$87.4&305.8 &86.6 &0.44 & 1&1\\
ZZ Gem  &343177896372048
 &4.66&2.92&3.14&1289.4&$-$4.49&66.8&9.61&0.15&51.6&250.5&10.1&0.18&   0&  2 \\
V758 Mon &305900810058616
 &5.03&3.74&2.94&2562.2&$-$5.36&49.9&10.41&0.25&$-$6.7&222.8&2.4&0.06& 0& 1\\
AQ And  &366243754095469
 &3.18&1.64&3.07&626.2&$-$4.28&$-$14.0&8.61&$-0$.25&$-$26.3&236.0&$-$19.6&0.08&0  &    1\\ 
AW Cyg  &212679878056696
 &3.56&1.90&3.12&1091.3&$-$5.18&$-$12.0&8.18&0.27&26.3&243.3&$-$8.2&0.08&  0 &   1 \\
EL Aur  &256672747465965
 &2.78&1.20&3.08&1021.3&$-$5.76& --  &9.28&0.12&--&--&--&--& --&\\
HK  Lyr  &209672462928414
 &3.04&1.55&3.05&958.9&$-$5.31&$-$5.0&8.01&0.31&27.4&256.9&$-$12.4&0.11& 0  &   1\\ 
LQ Cyg  &217315584007283
 &4.58&2.81&3.15&1590.2&$-$5.05&--     &8.67&0.01&--&--&--&--& --&\\
SY Per  &271405447366339
 &3.42&1.81&3.10&1034.5&$-$5.17&$-$1.0&9.27&0.02&13.9&223.7&$-$2.0&0.08& 0  &   5\\ 
TT Cyg  &20333717314002
 &3.16&1.79&2.99&658.6&$-$4.32&$-$4.9&8.10&0.08&$-$2.7&242.7&18.1&0.02&   0 &  1\\ 
TY Oph  &428419397614389
 &3.41&1.72&3.12&853.1&$-$4.81&$-$19.0&7.65&0.11&26.1&247.3&15.4&0.09&   0&   4 \\
V Oph  &433149585971175
 &3.07&1.42&3.11&717.2&$-$4.75&$-$28.2&7.68&0.32&23.1&262.5&10.5&0.12&  0 &  2 \\
V781 Sgr &406341551169915
 &3.08&1.49&3.09&1237.7&$-$5.89& --  &7.10&$-0$.01&--&--&--&--&-- &\\
SU And  &384828730421099
 &3.92&2.38&3.07&1373.7&$-$5.24&$-$6.0&8.90&$-0$.40&$-$2.3&256.8&0.3&0.09&0 &1\\
OR Cep  &530800890560157
 &5.15&2.65&3.20&899.3&$-$3.92& --   &8.80&0.14&--&--&--&--&-- &\\
CP Cas  &536229557425253
 &4.33&2.42&3.18&1363.2&$-$5.07& --   &9.12&0.20&--&--&--&--& --&\\
W Cas  &425942081532427
 &3.26&2.25&2.76&1478.3&$-$5.85&$-$40.7&9.23&$-0$.08&$-$35.1&247.9&10.9&0.12& 0  &   1\\ 
IRC +60041 &523048612026038
 &4.17&1.31&3.11&1090.7&$-$5.77& --     &9.01&0.03&--&--&--&--& --&\\
WW Cas  & 412902182864084736&3.61&2.10&3.05&1065.0&$-$4.99&$-$59.0&9.04&$-0$.06&$-$43.6&223.8&8.6&0.15& 0&     1\\ 
X Cas  &506349710470303
 &3.50&2.07&3.02&1229.7&$-$5.36&$-$55.0&9.19&$-0$.02&$-$25.6&214.1&6.7&0.17& 0 &    1 \\
BS Per  &451958622584743
 &4.02&2.25&3.15&1294.5&$-$5.16&$-$45.0&9.32&$-0$.15&$-$24.8&223.8&$-$2.7&0.10& 0  &   1 \\
R For  &511851181742149
 &3.79&1.05&3.15&622.0&$-$4.77&$-$1.0&8.53&$-0$.55&$-$6.9&228.1&16.5&0.05& 0 &   2 \\
V410 Per &435140973765206
 &3.61&2.10&3.05&1151.8&$-$5.15&8.0&9.30&$-0$.14&18.5&242.7&$-$7.7&0.07&  0 &   1 \\
V466 Per &443610786709962
 &2.16&0.62&3.07&614.5&$-$5.25&42.4&8.47&$-0$.13&25.6&251.9&$-$17.2&0.10&  0&    7\\
AC Per  &244490845826678
 &3.46&2.01&3.03&1146.0&$-$5.26& --   &9.36&$-0$.13&--&--&--&--&--     & 4 \\
IRC +40070 &229989043365962
 &6.03&2.81&2.94&1038.2&$-$4.33&$-$32.0&9.28&$-0$.16&$-$19.7&210.0&2.3&0.14& 0&3\\
GI Per  &179708170470777
 &2.81&1.04&3.15&1130.9&$-$6.08&9.0&9.41&$-0$.08&17.9&210.0&$-$11.2&0.14&   0 &  4\\ 
V1060 Tau &340506292710785
 &4.55&2.78&3.15&1220.7&$-$4.50&38.0&9.50&$-0$.34&32.9&249.5&26.9&0.13&  0 &   1 \\
TT Tau  &155132191506965
 &2.01&0.87&2.85&714.4&$-$5.55&16.0&9.04&$-0$.09&4.5&238.2&$-$17.9&0.02& 0 &    1 \\
IRC +20095 &34132316113066
 &4.38&2.58&3.16&1427.4&$-$5.04&48.0&9.73&$-0$.29&35.7&241.0&$-$8.8&0.12& 0   &  4 \\
TX Aur  &188467429855857
 &3.38&2.07&2.95&1364.0&$-$5.65&16.0&9.67&0.01&16.3&233.4&$-$18.3&0.05&  0 &   1\\ 
SY Eri  &321165900358034
 &3.62&2.19&3.02&1269.9&$-$5.31&6.6&9.38&$-0$.51&$-$37.2&237.2&$-$21.6&0.12&  0 &   1 \\
V431 Ori  &338764178384081
 &3.51&1.57&3.19&1310.2&$-$5.83&$-$11.0&9.58&$-0$.30&$-$36.7&238.8&$-$14.3&0.12& 0  &   4 \\
V438 Aur &344345412793951
 &3.96&2.33&3.10&1435.9&$-$5.35&18.0&9.76&0.01&15.5&230.3&$-$4.4&0.06&  0 &   1 \\
V1368 Ori &32413341695789
 &4.02&2.07&3.19&816.1&$-$4.29&$-$17.0&9.09&$-0$.20&$-$32.9&253.4&13.2&0.13& 0 &    5 \\
S Aur  & 182625449700126
 &4.52&1.49&3.03&1072.2&$-$5.63&3.0&9.40&0.01&$-$1.8&222.2&4.5&0.07&  0 &   4 \\
RT Ori  &333400935898511
 &3.02&1.58&3.02&992.7&$-$5.38&6.5&9.26&$-0$.20&$-$20.1&236.4&$-$8.4&0.06& 0 &    1\\ 
SZ Lep  &290911171472609
 &4.11&2.64&3.03&921.6&$-$4.15&38.4&8.89&$-0$.39&27.3&245.2&7.7&0.09&  0 &   1 \\
IRC +20115 &340421991092841
 &4.68&2.90&3.15&1424.8&$-$4.72&21.4&9.75&$-0$.08&5.4&216.2&8.4&0.10&   0&   3 \\
V1187 Ori &301542476532442
 &4.73&2.21&3.20&852.3&$-$4.25&--   &9.03&$-0$.25&--&--&--&--& --&\\
IRC +10094 &334038119636953
 &5.30&2.57&3.15&1211.9&$-$4.70& --   &9.50&$-0$.17&--&--&--&--&-- &\\
TU Tau  &342853210588476
 &2.48&1.28&2.89&1086.4&$-$6.01&$-$23.1&9.42&$-0$.01&$-$37.6&225.3&11.3&0.13&0  &    1\\ 
FU Aur  &344477171652866
 &2.97&1.76&2.90&914.3&$-$5.15&17.5&9.25&0.04&7.3&248.0&$-$2.6&0.06&  0 &   1 \\
V633 Mon  &308176483018212
 &6.04&3.34&3.16&1346.7&$-$4.14& --      &9.28&0.28&--&--&--&--& --&\\
GK Ori  &33257291731403
 &3.82&2.08&3.14&1302.9&$-$5.35&52.0&9.56&$-0$.05&18.2&210.8&15.7&0.13&0  &    5\\ 
IV CMa  &289901383429709
 &3.92&2.10&3.16&1781.4&$-$5.99&23.0&9.42&$-0$.51&$-$33.2&238.7&$-$15.5&0.11& 0 &    5 \\
V636 Mon &300319455389326
 &5.30&2.23&3.02&1196.0&$-$5.14&--&9.29&$-0$.18&--&--&--&--& --&\\
AB Gem &337236125227823
 &4.30&2.57&3.14&1586.0&$-$5.30&11.0&9.88&0.12&$-$11.4&233.0&$-$13.5&0.04& 0 &4\\
CR Gem  &335730866321553
 &3.13&1.29&3.17&875.8&$-$5.26&28.0&9.18&0.08&7.5&234.6&$-$8.8&0.03& 0&      5\\ 
RV Aur  &957998747501272
 &4.66&2.94&3.13&1279.8&$-$4.46&$-$51.0&9.56&0.36&$-$53.3&227.4&$-$7.1&0.17&    0&  1\\ 
IRC $-$10122  &300319455389326
 &5.05&2.83&3.22&1216.5&$-$4.38&  --    &9.29&$-0$.18&--&--&--&--& --&\\
GM CMa  &292638016331889
 &4.40&2.17&3.22&1097.2&$-$4.81& --    &9.04&$-0$.20&--&--&--&--& --&\\
VW Gem  &937136235919461
 &3.71&2.41&2.94&1466.2&$-$5.47&14.0&9.77&0.33&1.9&226.4&$-$3.8&0.04& 0 &    1\\ 
CZ Mon &312731419530803
 &4.09&2.41&3.12&1324.0&$-$5.08&27.0&9.51&0.02&6.9&261.9&6.1&0.12& 0&      4 \\
DF Mon  &312564046373241
 &4.62&2.83&3.15&1045.2&$-$4.11&77.0&9.24&0.02&46.8&224.2&$-$4.2&0.16&0  &      4 \\
GY Mon  &310210830355836
 &3.22&2.04&2.88&1400.8&$-$5.81&38.8&9.49&$-0$.01&$-$1.1&229.2&2.3&0.04& 0  &   1\\ 
CL Mon  &312988698379655
 &3.78&1.46&3.22&1349.5&$-$5.97&29.0&9.55&0.11&$-$15.6&203.9&$-$7.2&0.16& 0 &   2 \\
RV Mon  &312970309047785
 &3.40&1.74&3.12&593.7&$-$4.02&16.4&8.86&0.07&12.3&277.5&4.7&0.19& 0&      1\\ 
RY Mon  &305173496860705
 &2.37&0.53&3.17&924.9&$-$6.13&7.0&9.05&0.02&$-$32.6&233.5&1.9&0.10&  0&    5 \\
VX Gem  &316758325891653
 &4.16&2.69&3.04&1400.6&$-$5.00&32.9&9.62&0.30&1.4&218.2&$-$2.2&0.09&  0&   2 \\
V569 Pup &293034441813861
 &5.93&2.82&2.99&940.7&$-$4.05& --    &8.92& 0.01&--&--&--&--& --&\\
BE CMi  &313518604572557
 &5.10&3.07&3.20&1553.0&$-$4.68&44.0&9.61&0.32&10.3&232.3&$-$7.5&0.04&0 &     4\\ 
IRC $-$30097 &56001901319375
 &3.93&1.91&3.20&1133.7&$-$5.86&--&7.21&0.09&--&--&--&--& --&\\
GO Pup  &303678824189644
 &4.94&2.82&3.21&1737.8&$-$5.17&32.0&9.53&0.23&$-$8.1&243.3&13.4&0.04&0 &      4\\ 
V497 Pup &559726845466622
 &6.55&4.78&3.15&2399.2&$-$3.97&61.3&9.56&0.01&8.63&224.5&14.0&0.06&0  &3\\
RU Pup  &569900482247128
 &3.41&1.88&3.07&1345.7&$-$5.70&34.0&9.04&0.14&$-$10.8&228.4&$-$10.0&0.06& 0 &     5\\ 
AC Pup  &572213074745134
 &4.09&2.60&3.04&1196.0&$-$4.74&40.1&9.01&0.27&7.2&234.6&10.9&0.03&  0  &  1\\ 
IRC $-$30132  &565315882971833
 &5.53&2.65&3.10&1646.7&$-$5.34&--&8.43 &$-0$.02&--&--&--&--& --&\\
R Pyx  &564285939507045
 &4.51&2.51&3.20&1168.1&$-$4.63&25.4&8.77&0.21&10.8&245.2&2.5&0.05&0  &    2\\ 
IQ Hya  &565167544679535
 &5.60&2.86&3.15&1222.8&$-$4.43&$-$12.7&8.77&0.38&$-$26.4&276.1&20.8&0.20& 0 &     2\\ 
V Hya  &355062486203064
 &1.82&0.62&2.88&473.8&$-$4.87&$-$14.8&8.35&0.28&9.9&257.8&$-$13.1&0.09&  0 &   1\\ 
RU Vir  &3704116483406
 &3.23&1.53&3.13&1008.6&$-$5.36&1.5&8.14&0.95&$-$114.4&176.2&$-$38.1&0.43& 1 &   2 \\
SU Sco  &603034433082378
 &2.28&0.72&3.08&784.6&$-$5.67&$-$19.0&7.58&0.15&5.3&246.1&9.2&0.03&0    &  4 \\
V901 Sco  &602657194518541
 &5.24&2.87&3.21&1644.5&$-$5.00&1.0&6.72&0.17&$-$24.2&270.1&9.0&0.14&0   &   4 \\
V2309 Oph &405952506162438
 &3.22&1.48&3.14&1133.7&$-$5.65&$-$24.0&7.21&0.09&14.4&194.3&8.0&0.23& 1   &   4 \\
T Dra  &142197695202235
 &4.57&2.10&3.20&767.5&$-$4.12&$-$17.5&8.32&0.40&78.5&228.8&10.4&0.25&  0  & 2 \\ 
V4378 Sgr &40953052315004
 &4.25&2.38&3.18&898.1&$-$4.21&7.4&8.29&$-0$.09&47.5&257.9&$-$34.8&0.18&   0  & 7 \\ 
V1280 Sgr &406452409425722
 &5.85&2.57&2.90&1354.8&$-$5.19& --   &6.59&$-0$.09&--&--&--&--& --&\\
ES Ser  &415267255989212
 &3.15&1.44&3.13&835.2&$-$5.04&--&7.54&0.03&--&--&--&--& --&\\
IRC $-$20482 &409016800396156
 &3.78&1.83&3.19&1360.9&$-$5.65&--&7.01&$-0$.09&--&--&--&--& -- &\\
IRC +00351 &427635724136856
 &4.78&2.32&3.20&896.2&$-$4.24& --  &7.59&0.11&--&--&--&--& --&\\
SS Sgr  &409681389897537
 &4.25&2.55&3.13&1419.2&$-$5.08&2.1&6.98&$-0$.05&2.1 &256.8&3.5&0.06&0 &      1\\ 
DR Ser  &4285226383221
 &3.00&1.66&2.97&1369.9&$-$6.05&$-$14.9&7.30&0.10&38.0&252.3&$-$1.7&0.12&0 &     1 \\
NR Sgr  &407805733003885
 &4.37&2.42&3.19&976.4&$-$4.33& --   &7.39&$-0$.14&--&--&--&--&-- &\\
T Sct  &420398748917308
 &4.01&2.62&3.00&1560.7&$-$5.35&12.0&6.97&$-0$.10&3.8& 255.3&$-$15.9&0.06&0 &      1\\ 
UV Aql  &450677557103575
 &2.43&0.79&3.11&772.2&$-$5.54&22.9&7.83&0.09&$-$10.0&267.8&11.5&0.13&   0&   1\\ 
V1942 Sgr  &418397840471422
 &2.41&0.96&3.03&542.0&$-$4.68&$-$45.1&7.85&$-0$.10&40.1&231.7&$-$6.0&0.13&  0 &   1 \\
IRC +20406  &20196971053032
 &4.08&2.59&3.05&2010.6&$-$5.88&5.3&7.46&0.13&36.8&251.0&$-$1.60&0.12&  0 &   3 \\
V374 Aql &421525559352008
 &4.44&2.07&3.21&1606.6&$-$5.75&11.0&7.13&$-0$.23&8.6&250.1&$-$6.0&0.04&  0  &  4 \\
CU Vul &1.2749858576546
 &4.28&2.70&3.08&1654.9&$-$5.31&--   &7.61&$-0$.03&--&--&--&--& --&\\
AX Cyg  &207902517213906
 &3.19&1.42&3.15&926.5&$-$5.27&$-$6.0&8.21&0.15&$-$45.5&242.4&$-$4.9&0.14&0 &      1\\ 
V1583 Cyg  &203065799925555
 & 4.03&2.55&3.04&2292.8&$-$6.21&12.0&7.77&0.02&$-$86.5&242.2&$-$2.2&0.26& 0 &      4 \\
X Sge  & 1823613843707604736&3.24&1.76&3.04&1174.6&$-$5.55&26.0&7.81&$-0$.09&5.7&271.0&$-$12.9&0.14&0 &     4 \\
SV Cyg  &208289658264191
 &3.48&1.68&3.16&1208.8&$-$5.57&$-$8.0&8.28&0.19&$-$14.4&244.7&3.5&0.05& 0 &     1\\ 
AY Cyg  &207459149460516
 &4.35&2.54&3.16&1462.2&$-$5.12&16.0&8.15&0.13&4.8&266.7&$-$10.4&0.12&  0 &   5 \\
RY Cyg  &205895200685335
 &3.63&2.41&2.90&1113.2&$-$4.92& --     &8.09&0.05&--&--&--&--&   -- &  \\
RS Cyg  &206172026788756
 &2.09&0.85&2.91&729.4&$-$5.55&$-$50.0&8.19&0.05&37.6&206.9&$-$16.6&0.19&0  &      1\\ 
U Cyg  &208422136101664
 &3.02&1.20&3.16&752.2&$-$5.02& --   &8.29&0.11&--&--&--&--& -- &      \\
IRC +40436  &206278074683945
 &4.01&2.14&3.18&1179.6&$-$5.04&18.0&8.15&$-0$.02&15.4&268.8&12.2&0.14& 0&      4\\ 
V569 Cyg  &186286036149039
 &4.43&2.62&3.16&1238.0&$-$4.68&$-$26.0&8.09&$-0$.11&$-$40.4&216.0&6.2&0.17& 0&     4 \\
V1862 Cyg  &186901044542274
 &4.69&2.51&3.22&1466.2&$-$5.10&$-$8.0&8.10&$-0$.14&29.6&242.9&14.6&0.09&   0&   1 \\
DS Cyg  &216326215525576
 &4.26&2.47&3.15&1189.2&$-$4.75&$-$6.0&8.32&0.04&$-$22.6&248.6&$-$21.8&0.08&   0&   4\\ 
V1549 Cyg &216954078326394
 &6.81&2.65&2.12&660.7&$-$4.33& --   &8.38&0.06&--&--&--&--& --&\\
RV Aqr  &268990761341237
 &4.59&1.35&2.93&841.8&$-$5.35&$-$1.0&7.88&$-0$.39&14.2&263.8&17.6&0.12&0  &      4\\ 
YY Cyg  &19703564522407
 &4.54&2.57&3.20&1189.0&$-$4.61&$-$10.0&8.36&$-0$.08&$-$31.1&246.6&11.4&0.10& 0 &      5  \\
AX Cep &227217180994179
 &3.41&1.34&3.21&1404.9&$-$6.19&--    &8.83&0.36&--&--&--&--& --&\\
V1426 Cyg  &196583532914201
 &3.40&0.71&3.16&654.3&$-$5.21&$-$11.4&8.32&$-0$.08&8.2&241.5&11.7&0.03&0 &      3  \\ 
S Cep  &228471156825671
 &3.04&0.14&3.09&524.0&$-$5.37&$-$34.0&8.55&0.20&2.3&222.1&$-$18.1&0.08& 0&     1 \\
LU Cep  &221579651589698
 &4.10&2.55&3.07&1727.6&$-$5.57&$-$15.0&8.83&0.22&20.9&241.8&$-$6.5&0.07& 0 &     4\\
IRC +70172 & 2217977843882752128&5.01&3.05&3.19&2146.0&$-$5.42&--&9.11&0.39&--&--&--&--&-- &\\
V644 Cyg &197410520278021
 &6.91&3.46&2.79&1055.7&$-$3.87&--    &8.43&$-0$.07&--&--&--&--& --&\\
PQ Cep  &227393263915503
 &3.43&0.91&3.20&640.7&$-$4.93&  --   &8.57&0.19&--&--&--&--&-- &\\
V1398 Cyg &217405671804704
 &4.69&2.96&3.14&1597.0&$-$4.92& --    &8.59&0.04&--&--&--&--& --&\\
V413 Cyg  &217378966985691
 &3.65&1.94&3.13&1426.6&$-$5.70&$-$16.0&8.68&0.01&$-$4.7&245.9&11.6&0.03& 0&3\\
LW Cyg  &197981362663791
 &3.15&1.42&3.14&951.6&$-$5.34&$-$18.0&8.50&$-0$.02&0.9&237.2&9.0&0.01& 0 &    1\\ 
CT Lac  &197577701020797
 &4.58&2.40&3.21&2596.6&$-$6.40&$-$4.0&9.02&$-0$.24&49.6&242.0&$-$24.7&0.16& 0 &      1\\ 
V451 Cep  &200803231768394
 &4.44&2.4&3.20&1500.1&$-$5.27&$-$18.0&8.86&0.03&15.0&237.5&44.2&0.05&  0 &   1 \\
DG Cep  &220112796563072
 &2.79&1.60&2.88&712.3&$-$4.78&$-$30.0&8.59&0.05&$-$22.9&233.9&$-$9.6&0.08&  0 &   1 \\
IRC +60393 &201523271008621
 &3.97&1.92&3.21&1119.4&$-$5.12&$-$62.0&8.82&0.05&$-$29.2&207.4&5.4&0.17& 0 & 3\\
DS Cas  &201578518031916
 &4.44&2.52&3.19&1238.7&$-$4.76&--    &8.91&0.04&--&--&--&--& --&\\ 
&&&&&&&&&&&&&& &\\
R-hot &&&&&&&&&&&&& &&\\
HIP 2700 &280575916790125
 &7.59&6.89&2.49&785.0&$-0$.08&12.5&8.63&$-0$.48&37.5&236.3&$-$8.1&0.12&0 &3\\
HIP 5809&493219888490127
 &7.91&7.19&2.52&633.9&0.70&31.7&8.25&$-0$.56&3.1&186.4&$-$5.6&0.24& 1&3\\
HIP 18564&330487700813325
 & 7.39&6.56&2.61&670.6&0.05&$-$34.3&8.92&$-0$.31&$-$46.5&201.8&11.7&0.22&1 &3\\
HIP 28172&345174815216909
 & 7.73&7.20&2.34&814.8&$-0$.01&$-$56.3&9.15&0.09&$-$63.4&207.0&$-$4.5&0.24& 1&3\\
HIP 40374&554630835846382
 &8.96&8.28&2.48&2029.1&$-0$.77&204.2&9.15&0.02&63.7&81.6&$-$184.4&0.39& 2 &3\\
HIP 44172&713370948058513
 &7.30&6.61&2.49&615.2&0.16&25.2&8.80&0.42&1.9&255.0&29.0&0.09& 0 &3\\
HIP 44812&685257226010221
 &7.55&6.50&2.78&1366.8&$-$1.39 &19.7&9.29&0.89&8.7&211.9&$-$17.2&0.11&0 &3\\
HIP 48329&828468821685739
 &7.15&6.19&2.72&2024.3&$-$2.62&$-$6.7&9.63&1.53&$-$16.0&147.7&35.3&0.39& 1 &3\\
HIP 50994&535886193252966
 &7.27&6.60&2.47&708.4&$-0$.18&$-$2.3&8.24&0.09&$-$23.3&251.2&0.5&0.09& 0& 3\\
HIP 53810&353571713792282
 &6.38&5.7&2.49&252.8&1.17&5.5&8.32&0.15&$-$45.6&232.7&$-$16.1&0.15& 0&3\\
HIP 56405&379995816321967
 &8.12&7.44&2.48&761.9&0.52&11.6&8.39&0.67&15.3&244.3&12.9&0.06&0 &3\\
HIP 58786&168375468902166
 &8.15&7.47&2.48&1147.8&$-0$.35&$-$21.4&8.85&0.85&32.0&211.8&$-$10.2&0.15&0 &3\\
HIP 62944&396162262202678
 &4.78&4.1&2.48&127.8&1.05&5.7&8.33&0.15&$-$12.9&250.7&13.0&0.07&0 &3\\
HIP 63955&350599709810434
 &6.56&5.95&2.41&638.8&$-0$.66&$-$9.8&8.06&0.45&19.4&225.7&$-$3.8&0.09& 0&3\\
HIP 65320&37156754773495
 &7.56&6.85&2.51&420.9&1.24&$-$14.2&8.20&0.40&45.8&268.5&18.1&0.20&0 &3\\
HIP 66317&611296050066283
 &9.34&8.59 &2.55&2155.6&0.22& $-$6.40& 7.15&0.78&$-$13.2& 209.3&$-$24.9 &0.14&0 & 3\\
HIP 69089&151191841192363
 &6.92&6.36&2.36&655.2&$-0$.36&$-$20.2&8.38&0.60&$-$44.3&239.1&$-$5.0&0.14& 0&3\\
HIP 73955&172229697391808
 &8.09&7.51&2.39&1316.9&$-0$.70&$-$14.8&8.91&0.74&15.5&227.7&11.6&0.07&0 &3\\
HIP 74826&127519795651216
 &7.99&7.4&2.39&737.7&0.46&$-$95.2&8.07&0.64&78.3&124.9&$-$11.7&0.54& 1 &3\\
HIP 75745&599987978851385
 &8.86&8.16&2.50&1282.3&0.12&$-$32.8&7.26&0.25&19.8&233.6&9.3&0.07&  0&3\\
HIP 82184&456645764627615
 & 7.14&6.46&2.48&556.8&0.22&$-$30.8&8.01&0.35&$-$14.6&226.1&$-$31.7&0.08& 0 &3\\
HIP 84266&135415103464298
 &5.81&5.27&2.33&286.7&0.32&$-$12.2&8.25&0.19&$-$52.7&221.7&1.3&0.18& 0&3\\
HIP 85117&597453955016864
 &6.11&5.24&2.64&515.1&$-0$.67&$-$80.5&7.83&0.02&71.9&229.1&$-$2.3&0.22& 0 &3\\
HIP 85750&437648488622391
 &6.15&5.10&2.79&556.8 &$-0$.84&$-$22.0&7.23&0.44&16.0&238.1&$-$36.0&0.04&0  &3\\
HIP 86927&454945304611911
 &6.59&5.95&2.45&423.3&0.27&$-$47.4&8.05&0.18&27.6&219.2&$-$8.0&0.13& 0 &3\\
HIP 88887&448237850320077
 &6.24&5.09&2.86&423.3 &$-0$.19 &$-$15.3&8.01&0.12&$-$2.3&230.6&1.5&0.05&0 &3\\
HIP 89239&449857043112816
 &8.14&7.36&2.58&1330.7&$-0$.68&10.4&7.43&0.37&18.6&261.1&22.5&0.11&0 &3\\
HIP 90199&44768059721116
 &9.53&9.06&2.27&558.4&2.60&$-$32.2&7.89&0.10&$-$6.1&203.7&$-$38.6&0.16& 1&3\\
HIP 102726&685844900800262
 &8.02&7.32&2.50&1039.4&$-0$.26&56.7&7.58&$-0$.56&40.5&241.0&$-$21.5&0.19&0 &3\\
HIP 105212&268613783471714
 &7.66&7.01&2.46&446.6&1.22&17.3&8.10&$-0$.21&41.5&201.3&$-$38.1&0.23& 1&3\\
HIP 105241 &269289055409865
 &7.60&6.88&2.52&1110.4&$-0$.83&$-$52.8&7.83&$-0$.54&42.5&194.2&$-$6.2&0.26&1 &3\\
HIP 114452&194140771982304
 &7.50&6.90&2.40&837.6&$-0$.31&$-$41.2&8.59&$-0$.16&44.5&219.7&$-$1.4&0.12& 0&3\\
HIP 114509&23866511395564
 &7.24&6.58&2.45&784.9&$-0$.43&42.1&8.11&$-0$.69&45.5&188.2&$-$85.0&0.32& 1&3\\
HIP 117467&274483193007067
 &6.17&5.45&2.51&291.4&0.65&$-$20.5&8.36&$-0$.20&46.5&227.7&16.4&0.06& 0&3\\
& & & & & & & & & & & && & &\\
O-rich &&&&&&&&&&&&&& &\\
YZ Peg     &19020039374988
 &5.37&4.00&2.99&259.6&$-0$.08&--   &8.34&$-0$.06&--&--&--&--& --&\\
V1828 Cyg  &206315183200366
 &4.39&2.29&2.94&933.5&$-$4.62&--   &8.19&$-0$.01&--&--&--&--& --&\\
WX Ser     &121076901466322
 &3.81&1.84&2.94&753.9&$-$4.60&$-$7.0 &7.95& 0.63&15.2&276.8&1.5&0.18&0 &4 \\ 
V1012 Her  &138342793691377
 &4.79&3.65&2.85&1913.5&$-$4.91&--  &7.92& 1.44&--&--&--&--& --&\\
RX Oph     &444060833563479
 &4.35&2.94&3.01&852.5&$-$3.70&$-$65.0&7.66 &0.43 &61.2&208.1&9.9&0.23&1 &8\\
RS Vir     &366910315409402
 &2.45&1.37&2.81&613.9&$-$4.77&$-$19.2&8.01&0.54 &22.8&260.9&7.2&0.12& 0&1 \\ 
IRC +30074 &165604872260654
 &3.53&2.16&2.98&587.0&$-$3.70&--   &8.89&$-0$.13&--&--&--&--&-- &\\
IRC +10119 &332558458736027
 &4.72&3.23&3.04&622.6&$-$2.69&--&8.91&0.01&--&-- &--&--& --&\\
V515 Pup   &55958627977733
 &3.81&2.21&3.09&557.0&$-$3.43&--   &8.55&0.02 &--&--&--&--&-- &\\
V1617 Cyg  &21730579491786
 &3.00&1.77&2.91&655.0&$-$4.41&--   &8.43&0.01&--&--&--&--& --&\\

\end{longtable}

\tablefoot{P: membership probability higher than 80\%  for thin disc (0), thick disc (1) or halo (2), see text.
  References for V$_{\rm{rad}}$: (1) \citet{gon06}; (2) \citet{men06}; (3) \citet{gai18}; (4) \citet{duf95}; (5) \citet{kha07};
  (6) \citet{cat71}; (7) \citet{kun17}; (8) \citet{fea00}. }
\end{landscape}
}

\bibliographystyle{aa} % style aa.bst
\bibliography{aa.bib} % your references Yourfile.bib
\end{document}